\documentclass[onecolumn]{aastex61}
\usepackage[]{hyperref}
\usepackage{float}
\usepackage{color}
\usepackage{graphicx}
\usepackage{amsmath}
\usepackage{natbib}
\newcommand{\cgs}[1]{erg cm$^{-2}$ s$^{-1}$ \AA$^{-1}$}
\newcommand{\prim}{$^{\prime}$}

\begin{document}

\title{The Near-Ultraviolet Continuum Radiation in the Impulsive Phase of HF/GF-Type \lowercase{d}M\lowercase{e} Flares I:  Data}

\author{Adam F. Kowalski}
\affil{Department of Astrophysical and Planetary Sciences, University of Colorado Boulder, 2000 Colorado Ave, Boulder, CO 80305, USA.}
\affil{National Solar Observatory, University of Colorado Boulder, 3665 Discovery Drive, Boulder, CO 80303, USA.}
\affil{Laboratory for Atmospheric and Space Physics, University of Colorado Boulder, 3665 Discovery Drive, Boulder, CO 80303, USA.}
\email{Adam.F.Kowalski@colorado.edu}
\author{John P. Wisniewski}
\affil{Homer L. Dodge Department of Physics and Astronomy, University of Oklahoma, 440 W. Brooks Street, Norman, OK 73019, USA.}
\author{Suzanne L. Hawley}
\affil{University of Washington Department of Astronomy, 3910 15th Ave NE, Seattle, WA 98195, USA.}
\author{Rachel A. Osten}
\affil{Space Telescope Science Institute, 3700 San Martin Drive, Baltimore, MD 21218, USA.}
\affil{Center for Astrophysical Sciences, Johns Hopkins University, Baltimore, MD 21218}
\author{Alexander Brown}
\affil{Center for Astrophysics and Space Astronomy, University of Colorado, 389 UCB, Boulder, CO 80309-0389, USA. }
\author{Cecilia Fari\~{n}a}
\affil{Isaac Newton Group of Telescopes, E-38700 Santa Cruz de La Palma, La Palma, Spain.}
\affil{Instituto de Astrof\'{i}sica de Canarias (IAC) and Universidad de La Laguna, Dpto. Astrof\'{i}sica, Spain.}
\author{Jeff A. Valenti}
\affil{Space Telescope Science Institute, 3700 San Martin Drive, Baltimore, MD 21218, USA.}
\author{Stephen Brown}
\affil{School of Physics and Astronomy, Kelvin Building, University of Glasgow, G12 8QQ, Scotland.}
\author{Manolis Xilouris}
\affil{Institute for Astronomy, Astrophysics, Space Applications \& Remote Sensing, National Observatory of Athens, P. Penteli, GR-15236 Athens, Greece.}
\author{Sarah J. Schmidt}
\affil{Leibniz-Institut f\"{u}r Astrophysik, An der Sternwarte 16 14482 Potsdam, Germany.}
\author{Christopher Johns-Krull}
\affil{Department of Physics \& Astronomy, Rice University, 6100 Main St. MS-108, Houston, TX 77005, USA.}

\begin{abstract}
We present NUV flare spectra from the \emph{Hubble Space Telescope}/Cosmic Origins Spectrograph during two moderate-amplitude $U$-band flares on the dM4e star GJ 1243.  
These spectra are some of the first accurately flux-calibrated, NUV flare spectra obtained over the impulsive phase in M dwarf flares.  We observed these flares  
with a fleet of nine ground-based telescopes simultaneously, which provided broadband photometry and low-resolution spectra at the Balmer jump.  
A broadband continuum increase occurred with a signal-to-noise $>$ 20 in the HST spectra, while numerous Fe II lines and the Mg II lines also increased but with smaller flux enhancements compared to the continuum radiation.  These two events produced the most prominent Balmer line radiation and the largest Balmer jumps that have been observed to date in dMe flare spectra.  
A $T=9000$ K blackbody under-estimates the NUV continuum flare flux by a factor of two and is a poor approximation to the white-light in these types of flare events.  Instead, our data suggest that the peak of the specific continuum flux density is constrained to $U$-band wavelengths near the Balmer series limit. 
A radiative-hydrodynamic simulation of a very high energy deposition rate averaged over times of impulsive heating and cooling better explains the $\lambda>2500$ \AA\ flare continuum properties.  These two events sample only one end of the empirical color-color distribution for dMe flares, and more time-resolved flare spectra in the NUV, $U$-band, and optical from $2000-4200$ \AA\ are needed during more impulsive and/or more energetic flares.

\end{abstract}


\section{Introduction \& Motivation}
The near-ultraviolet (NUV) spectral range ($\lambda=2000-4000$ \AA) in stellar flares is critical for understanding the underlying physics of the impulsive release of magnetic energy in flares, the explosive hydrodynamic response of the stellar atmosphere, and the effects on ozone chemistry and surface biology of (potentially) habitable worlds around M dwarfs.   However, the spectral characteristics of the NUV wavelength range are a critical missing piece in the observational 
picture of stellar flares.  Stellar flares are often observed in the NUV with $U$-band ($3200-4000$ \AA) photometry, but spectral observations at $\lambda < 3500$ \AA\ are rare.  While part of the NUV was sampled during the Great Flare of AD Leo \citep{HP91}, these observations only
covered part of the gradual decay phase\footnote{The International Ultraviolet Explorer (IUE) LWP spectra from  $1900-3100$ \AA\ observations of the Great Flare of AD Leo started at 1200 s in Figure 1 of \citet{HP91}.}.  Though nearly half of white-light flare radiation is emitted in the NUV where the continuum distribution is thought to peak \citep[][]{HP91}, there is a lack of time-resolved impulsive phase flare spectra in the NUV.  \citet{Robinson1993, Robinson1995} present the only flux-calibrated spectrum (also from IUE) of a flare impulsive phase, though this spectrum has a 20-minute integration time and includes a significantly long time of gradual decay phase radiation.

With accurately flux-calibrated spectra, the flare continuum peak ($\lambda_{\rm{flare,peak}}$) in the NUV ostensibly provides a high-cadence characterization of the temperature evolution in the lower atmosphere through Wien's Law.  The temperature evolution is difficult to constrain from optical spectra\footnote{In the optical, the systematic errors on color temperature measurements are as large as 2000 K for low-amplitude flares and in the gradual decay phase \citep{Kowalski2013}. } at $\lambda > 4000$ \AA\ where there is much lower contrast against the non-flaring red photosphere of M dwarfs.  Time-resolved impulsive phase spectra in the NUV have much larger flare contrast and would robustly constrain  $\lambda_{\rm{flare,peak}}$ and its time evolution without being affected by uncertainties from the subtraction of the pre-flare spectrum. 

Broadband photometry in the FUV, NUV, and the Johnson $U, B, V$, and $R$ filters during the Great Flare on AD Leo \citep{HP91} suggest that the continuum flux distribution (with specific flux density units of \AA$^{-1}$) exhibits a value of $\lambda_{\rm{flare,peak}}$ within the $U$ band and decreases toward FUV wavelengths \citep{HF92} like a $T\sim9000-10000$  K blackbody.  Other moderate-amplitude flares on AD Leo also exhibit the general distribution of a $T\sim9000$ K blackbody \citep{Hawley2003}.  However, a giant flare observed in the FUV and NUV GALEX bandpasses have suggested temperatures much higher with $T>50,000$ K \citep{Robinson2005}.  The FUV evolves faster than the NUV \citep{Hawley2003, Welsh2006} which may indicate that a homogeneous flare source does not explain all of the FUV and NUV radiation \citep[e.g., see Appendix of ][]{K18, Kowalski2017B}.  Thus, detailed spectra of the NUV will help distinguish between lines and continua and constrain the broadband spectral shape.  
Furthermore, broadband photometry can result in
 degeneracies among emission models \citep{Allred2006}, so spectral characterization at high time-cadence using a combination of space-based and ground-based observations that cover $\lambda=2000-4800$ \AA\ are important for a comprehensive understanding of the heating at high column mass in stellar flares.   
 
 Spectral observations from the ground in the $U$-band indicate that blackbody radiation alone does not explain white-light radiation. A Balmer continuum component is necessary to explain the jump in flux \citep{Kowalski2010}, but the jump is much smaller than expected from a source emitting hydrogen recombination at $T\sim10,000$ K over low continuum optical depth \citep[][hereafter, K13]{Kowalski2013}.  The spectral shape in the impulsive phase at $\lambda=3400-3650$ \AA\ roughly exhibits the same color temperature of $\sim10,000-12,000$ K as inferred in the blue-optical ($4000-4800$ \AA) wavelength regime (K13).  
 
 There are several spectral observations of flares below the atmospheric cutoff, but they are not readily comparable to models. The flux calibration accuracy of the echelle flare spectra of YZ CMi from HST/STIS \citep{Hawley2007} has not been assessed (STIS ISR 1998-18) for the purpose of constraining the continuum shape and comparing to radiative-hydrodynamic (RHD) flare model predictions.  Moreover, no ground-based observations of the Balmer jump wavelength region ($3600-4000$ \AA) were available for these flares; the Balmer jump region and blue-optical wavelength region are essential for a correct interpretation of the NUV continuum spectrum as either optically-thick (hot blackbody-like in the optical) or optically thin Balmer continuum radiation (cool blackbody-like in the optical).  
  \citet{Wargelin2017} present Swift UV grism spectra during a flare on Proxima Centauri, but the data have not been flux-calibrated, and the contribution from second order light has not been assessed for this purpose \citep{Kuin2015}. 

 A blackbody-like spectrum with a color temperature of $T \sim10,000$ K in the impulsive phase of some dMe flares means that there is significant heating at high column mass \citep[$m\gtrsim 0.01$ g cm$^{-2}$;][]{K18}, which is not possible to reproduce in RHD simulations of low-to-moderately high flux electron beam energy deposition rates \citep{Allred2006}.  RHD models with very high beam energy deposition rates \citep{Kowalski2015} suggest that the broadband appearance of a hot blackbody can be explained by hydrogen recombination emissivity that escapes over regions of the atmosphere at $T\sim10,000$ K with wavelength dependent continuum optical depth of $\tau_{\lambda, \rm{continuum}}$  between $\sim0.4 - 5$.  These models also predict 
 evolution of  $\lambda_{\rm{flare,peak}}$ due to the evolving continuum optical depths in a chromospheric condensation that cools from $T\gtrsim50,000$ K to $T\sim10,000-12,000$ K over several seconds \citep[][hereafter, K16]{Kowalski2015, Kowalski2016}. Thus, the location of the $\lambda_{\rm{flare,peak}}$ may be more sensitive to continuum optical depth than flare temperature.  Most models with electron beam heating produce 10,000 K material, but 10,000 K material with large continuum optical depth requires extreme heating scenarios \citep[see also][]{Cram1982, Houdebine1992,Christian2003} that challenge how the standard solar flare paradigm applies to dMe flares, which can be much more energetic.  Accurately flux calibrated measurements of $\lambda_{\rm{peak}}$ in low-to-moderate-energy dMe flares would provide unambiguous comparisons between the heating of high column mass in solar and dMe flares; however, these observations are the most difficult to obtain due to the lack of accessibility to the $\lambda \sim 2000-3500$ \AA\ wavelength region except with the largest ground-based telescopes or from space.  
  
K13 classified a sample of dMe flares with NUV and optical spectra ($\lambda=3420 - 7000$ \AA) and $U$-band photometry according to the impulsiveness, which is the $U$-band peak flux enhancement (minus 1) divided by the full-width-at-half-maximum ($t_{1/2}$; in minutes) of the light curve.  The impulsive flare (IF) events are those with fast evolution to a bright peak amplitude, whereas the gradual flare (GF) events are those that exhibited longer rise times to the peaks, which sometimes result from a superposition of several relatively low-amplitude fast flares (e.g., the GF1 event in K13) and/or a gradually rising continuum flux (e.g., the GF1 event in K13; the GF2 event in K16).  The IF-type events are distinguished for their small Balmer jump ratios and small ratios of the H$\gamma$ line-integrated flux divided by the 4170 \AA\ (blue) continuum flux.  GF-type events exhibit larger Balmer jump ratios and H$\gamma$-to-blue continuum ratios.  Between the IF and GF events are events with intermediate values of the Balmer jump ratio and line-to-continuum ratio:  these are the ``hybrid'' flare (HF) events, which also show striking evidence (c.f. Figure 8 of K13) for hot ($9000-12,000$ K) blackbody-like radiation at $\lambda>4000$ \AA\ like most IF-type events in their spectroscopic sample\footnote{K16 analyzed another large sample of dMe flares with narrowband continuum photometry and found that some IF-type events, which make up most of the events in a sample that is limited by signal-to-noise at flare peak, exhibit larger Balmer jump ratios.}, but the amplitudes are rather low and require further confirmation for the presence of this spectral phenomenon.  The physical underpinning of the IF/HF/GF type classification (which, like spectral types, is a continuous classification scheme) of dMe flares and the relationship between the continuum color temperature and line-to-continuum ratios has not yet been solved, but NUV spectra at $\lambda<3500$ \AA\ for each type of flare would help establish which types show the dominant $T \sim 10^4$ K component that is so difficult to reproduce in RHD flare simulations.   

  In this paper, we describe the first time-resolved, accurately flux-calibrated flare spectra during two HF/GF-type dMe flare events in the NUV using spectral data from the Hubble Space Telescope (HST)/Cosmic Origins Spectrograph (COS) with simultaneous spectra at the Balmer jump and optical photometry. 
This is the first paper in a series and focuses on the presentation of the data and a comparison to existing radiative-hydrodynamic flare models.  In Section \ref{sec:data}, we describe the data reduction of the ground-based and HST observations; in Section \ref{sec:photanalysis}, we describe the derived properties from the broadband flare light curves; in Section \ref{sec:bj}, we present the Balmer jump properties and optical flare spectra;  in Section \ref{sec:discussion1}, we discuss the flare colors in the context of the events in K13 and K16; in Section \ref{sec:cosanalysis}, we present the HST/COS spectral characteristics and the time-evolution of the continuum and emission lines; in Section \ref{sec:combined}, we combine the HST data with the Balmer jump spectra to compare to radiative-hydrodynamic models in the literature;  in Section \ref{sec:discussion2}, we discuss the implications for heating at high column mass;  in Section \ref{sec:discussion3}, we discuss the significance of these new flare observations for photochemical modeling of planets in dM habitable zones; and in Section \ref{sec:conclusions}, we present a summary of what we have learned about dMe flares from this study.  In Paper II of this series, we will present new radiative-hydrodynamic modeling avenues of heating at high column mass that have been motivated by our new NUV constraints on the continuum radiation.

\section{Data} \label{sec:data}

On 2014 Sep 01, we monitored the dM4e star GJ 1243 with nine ground-based telescopes and the HST/COS for eight orbits (HST GO 13323, Dataset LCD201010) over an elapsed time of 11 hr and 15 min.  The datasets are summarized in the observing log in Table \ref{table:obslog}.
We chose this relatively faint ($V=12.8$) dMe star because at that time conservative bright object limits were in place for COS and STIS that were not employed for previous studies
of brighter flare stars (e.g., AD Leo, $V=9.5$).  After Servicing Mission 4 there was increased concern about stochastically occurring large amplitude stellar flares, which could cause detector shut down and potentially harm the photon counting detectors of STIS and COS. This necessitated a careful target selection for the proposed science.  Since 
2014, the bright limits for observing flare stars with HST have been set according to ISR STIS 2017-02 and ISR COS 2017-01.  

GJ 1243 has been monitored extensively by the \emph{Kepler} satellite, providing a robust white-light flare rate over eleven months of one-minute cadence data \citep{Hawley2014, Davenport2014, Silverberg2016}.  In the 5.6 hours of monitoring with COS, we were able to justify to the HST TAC that the robust flare rate from \emph{Kepler} guaranteed two moderate-sized flares and many smaller events would occur.  In this paper, we describe the data and analysis of the two moderate-amplitude flares with $\Delta U \sim -1.5$ mag (at peak) that occurred towards the end of the target visibility in the second (LCD201ADQ) and sixth orbits (LCD201DAQ) of HST.

\begin{deluxetable}{lccc}
\rotate
\tabletypesize{\scriptsize}
\tablewidth{0pt}
\tablecaption{Observation Log}
\tablehead{
\colhead{Telescope} &
\colhead{Instrument} &
\colhead{UT} &
\colhead{Wavelength range [\AA] / Filters}}
\startdata 
HST & COS & 2014-08-31 23:14 - 2014-09-01 10:29 & 2444 - 2841 \\
ARC 3.5m at APO &  DIS  &  2:03 - 10:54 &  3400 - 7500 \\
4.2m WHT at La Palma & ISIS &  2014-08-31 21:27 - 2014-09-01 4:05  & 3400 - 8000 \\
Keck 10m & LRIS &  5:06 - 10:20 & 3120 - 5390 \\
Harlan J. Smith 2.7m at McDonald & VIRUS-P & 2:11 - 9:35 & 3400 - 6850 \\
\hline
Otto Struve 2.1m at McDonald & 2-channel P45J photometer &  2:22 - 8:24 & $U$ \\ 
Aristarchos 2.3m at Helmos Observatory & RISE2 & 2014-08-31 22:05:37 -  2014-09-01 1:20  & $V+R$ \\
INT 2.5m at La Palma & WFC   & 2014-08-31 21:54 - 2014-09-01 4:16   & Str\"omgren $u$ \\
0.8m at McDonald & PFC & 5:24 - 7:59 & $BVRI$ \\
ARCSAT 0.5m at APO & Flarecam & 2:21 - 10:30 & SDSS $gri$ \\
\enddata 
\tablecomments{Unless noted, UT times correspond to times on 2014-09-01. }
\end{deluxetable}\label{table:obslog}

\subsection{Hubble Space Telescope/COS Spectra} \label{sec:cosdata}
Existing flare literature describes $U$-band measurements as NUV, whereas space observatories reserve the term NUV for shorter wavelengths not visible from the ground. In this paper, we will use ``$U$ band'' to describe ground-based (photometry or spectral) measurements at wavelengths longer than the atmospheric cutoff and ``NUV" to describe space-based (HST/COS) measurements at wavelengths shorter than the atmospheric cutoff.

	We employed the G230L grating ($\lambda_{\rm{cen}}=2635$ \AA) with COS on HST, resulting in wavelength coverage in the NUV from $\lambda \sim 2444$ \AA\ to $\lambda \sim 2841$ \AA\  (the NUVB stripe) and a dispersion of 0.39 \AA\ pixel$^{-1}$ ($42 - 47$ km s$^{-1}$ pixel$^{-1}$).  The target was acquired and centered in the Primary Science Aperture using the ACQ/PEAKXD and ACQ/PEAKD modes (with $\lambda_{\rm{cen}}=3360$ \AA\ for the acquisition). We used the standard Space Telescope Science Institute Python data extraction and reduction routines (\emph{splittag} and \emph{x1dcorr}).
	We extracted NUVB/TIME-TAG spectra at 5~s intervals and integrated over this wavelength range of COS/G230L for broadband light curve analysis in Section \ref{sec:photanalysis}.   

	At $2444-2509$ \AA, there is vignetting making this spectral range unsuitable for continuum shape characterization.  
	We extracted TIME-TAG spectra for the $2510-2841$ \AA\ spectral range for the 60~s exposure times corresponding to the spectra from the William Herschel Telescope (WHT; Section \ref{sec:wht}) and from the Apache Point Observatory (APO; Section \ref{sec:apodata}) for detailed spectral analysis in Section \ref{sec:cosanalysis}.  
        We employed the standard reference files for wavelength and flux calibration of the HST/COS data, but we used an aperture of $y_{\rm{cen}} \pm 15$ pixels without an aperture correction (the standard 
        aperture size is $\pm 57$ pixels).  This reduced aperture results in improved S/N for faint objects like M dwarfs in the NUV (ISR COS 2017-03).  However, the smaller aperture decreases the HST/COS fluxes of GJ 1243 by $10-15$\% in non-flaring and flaring times (consistent with the findings of ISR COS 2017-03);  this minor loss of light does not affect our calculation of the continuum flux ratios (Section \ref{sec:cos_peak_analy}) within the COS range by more than $\sim2$\%.
 Since the flare spectra exhibit a low count rate, the smaller aperture with slightly better S/N is important for identifying continuum regions that are free of faint emission lines and for calculating the line fluxes of the relatively faint Fe II emission lines.    In Section \ref{sec:combined}, we apply a wavelength-independent aperture correction of 14\% to the flare spectra for comparison to the absolute flux calibration of the Balmer jump and optical spectra from the ground-based observatories.  The wavelength solution from the pipeline was manually adjusted by $+2.4$ \AA; this shift is consistent with the quoted accuracy of COS/G230L.  We also discovered an unexplained jump in the wavelength solution by $\sim1$ pixel at approximately halfway through each orbit; this was not obviously related to any stellar activity or other known instrumental effects\footnote{STScI help desk, private communication.}.  Our results are not affected by this wavelength calibration uncertainty.  

\begin{figure}
\includegraphics[scale=0.45]{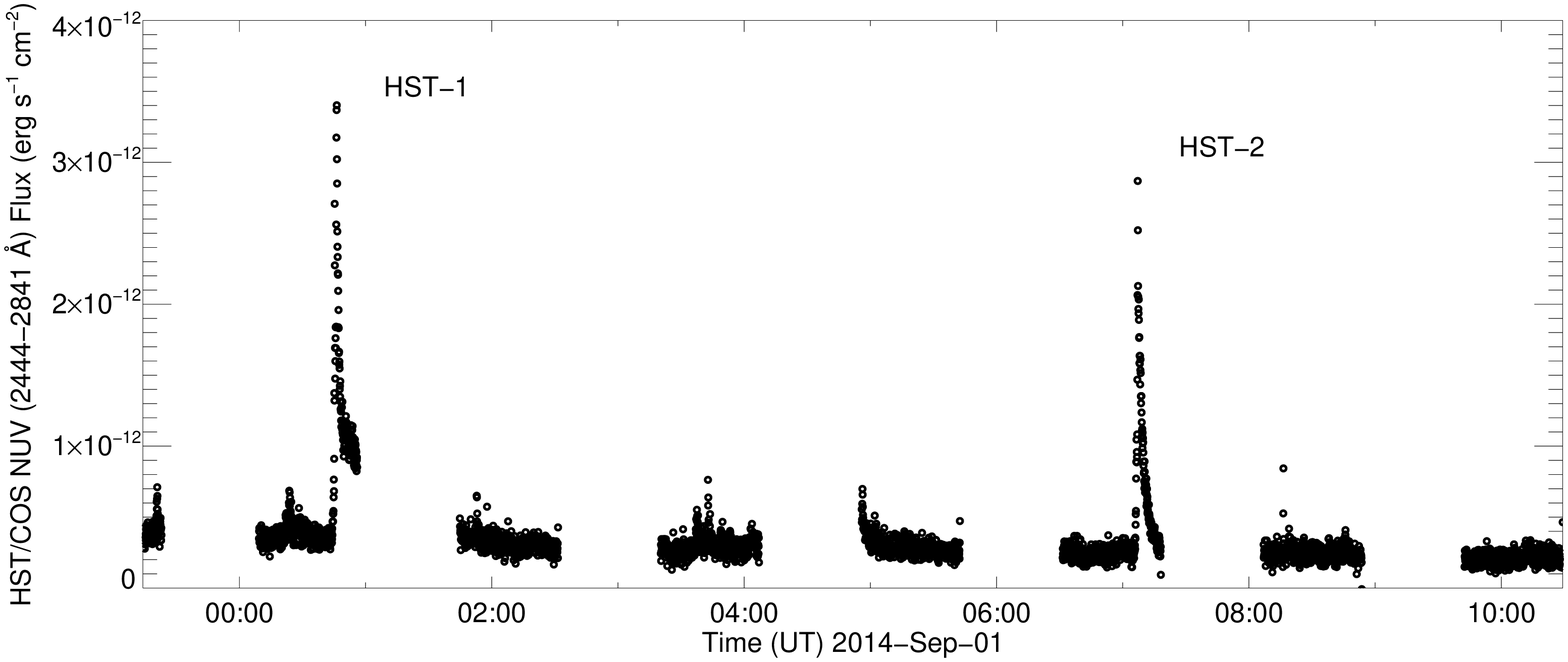}
\caption{Light curve of GJ 1243 for eight orbits of the HST/COS, showing the G230L specific flux density integrated over wavelength from $2444-2841$ \AA\ in 5 second time bins.  Two moderate-sized flares are indicated as HST-1 and HST-2 and are discussed throughout the text.   The target visibility in each orbit has a duration of 2800~s, and the spectral observations in the first orbit are shorter due to the target acquisition.  }
\label{fig:hstlc}
\end{figure}

The NUV light curve of GJ 1243 at 5~s time-binning is shown in Figure \ref{fig:hstlc}.  Two moderate-sized flares 
occurred in the HST monitoring at the ends of second and sixth orbits, and many smaller events also occurred. 
 The flare peaking at 00:46 UT on 2014 Sep 01 (hereafter, HST-1) and the flare peaking at 7:07 UT on 2014 Sep 01 (hereafter, HST-2) are analyzed 
in detail with the coordinated ground-based observations from Mauna Kea (Keck I/LRIS), the Apache Point
Observatory (ARC 3.5-m/DIS, ARCSAT 0.5-m/Flarecam), the McDonald Observatory
(Harlan J. Smith 2.7-m/VIRUS-P, Otto Struve 2.1-m/photometer, 0.8-m/PFC), the Isaac
Newton Group of Telescopes at La Palma (4.2-m WHT/ISIS, 2.5-m INT/WFC), and the
Helmos Observatory (Aristarchos 2.3-m/RISE2).

\clearpage
\clearpage

\subsection{Spectra from the ARC 3.5m at the APO} \label{sec:apodata}
We employed a 5\arcsec\ slit oriented at the parallactic angle for observations with the Dual-Imaging Spectrograph (DIS) on the ARC 3.5-m at APO.  The spectral resolution, measured\footnote{Through a wide slit that is much larger than the seeing, the actual spectral resolution from a point source can be significantly higher than indicated by the arc lamp lines.} from the quiescent hydrogen Balmer $\delta$ and $\alpha$ lines, is R$\sim$590 in the blue and $R\sim780$ in the red. The exposure times are 60~s, and the spectrophotometric standard stars Feige 110 and PG 1708$+$602 were used to flux calibrate the data.  
The data reduction and flux calibration were performed using IRAF, and the specific procedures applied for low-resolution flare spectroscopy are described in K13 and K16.  The quiescent spectra were scaled to the $B$-band mag (14.47) and the $V$-band mag (12.83) of GJ 1243 from \citet{Reid2004}, using the bandpass zeropoints from \citet{Bessel2013}.  
From all ground-based spectra  (and also for spectra in Section \ref{sec:wht}), we calculate the quantities from K13 and K16 from the flare-only (pre-flare subtracted excess) specific flux density at Earth ([erg cm$^{-2}$ s$^{-1}$ \AA$^{-1}$]), which is hereafter just denoted as ``flare-only flux'' with a prime symbol (e.g., $F_{\lambda}^{\prime}$, following \citet{Kowalski2012SoPh}; note that prime symbols were not used in the notation of K13 or K16).  Specifically, the calculated quantities are the following:
 the average flare-only flux in spectral windows that contain mostly continuum radiation (C3615\prim, C4170\prim, C6010\prim), the color temperature at blue optical wavelengths at $\lambda=4000-4800$ \AA\ ($T_{\rm{BB}}$), the Balmer jump ratio (C3615\prim/C4170\prim), the blue-to-red continuum flux ratio (C4170\prim/C6010\prim), the emission line fluxes (continuum subtracted) integrated over the line wavelengths\footnote{Corresponding to the 1.5\arcsec\ slit line windows in Table 3 of K13.}, and the line-to-continuum ratio of the H$\gamma$ line-integrated flux to C4170\prim\ (H$\gamma$/C4170\prim).  The HST-2 event occurred during these spectral observations.  
 
\subsection{Spectra from the WHT 4.2m} \label{sec:wht}
Data were obtained in the red and blue arms of Intermediate dispersion Spectrograph and Imaging System (ISIS) at the WHT with
gratings R158B and R316R.  The seeing varied from 1.1-1.8\arcsec\ through the night, and a 2\arcsec\
slit was employed.  The slit was repositioned to the parallactic angle
several times throughout the night (between HST orbital visibility periods).  Exposure
times were 10 sec for the red and 60 sec for the blue with a readout
of about 6 sec.  Standard IRAF
procedures were used to reduce the data.  Observations of GJ 1243
were obtained from 2014-08-31 21:27 UT to 2014-09-01 4:05 UT; the HST-1 event occurred at 00:46 UT through an airmass of sec $z = 1.2$, where $z$ is the zenith angle.
The wavelength solution was obtained with a CuNe$+$CuAr lamp at the
beginning of the night using a 0.7\arcsec\ slit.  The dispersion is
3.25 \AA\ in the blue and 1.8 \AA\ in the red.  The spectral resolution (measured from quiescent spectrum Balmer $\delta$ and $\alpha$ emission lines)
is $\sim480$ in the blue and $\sim1340$ in the red.
We used a spectrophotometric standard star PG1708$+$602 obtained at an
airmass of 1.2 at 2014-Aug-31 20:32 UT to flux calibrate the data.   Several more
observations of standard
stars
were obtained through the night to assess the residual extinction
compared to the standard extinction curve from King (1985);  large
flux variations
were found from star to star at $\lambda \gtrsim 8000$ \AA.    In the
blue, the usable wavelength range is $3400 - 5320$ \AA\ and in the red
the usable range is $5500 - 8000$ \AA.

\subsubsection{Calculation of flare-only flux}
We use the spectrum scaling algorithm of K13 and K16 to correct for 
variable slit loss due to seeing changes through the night, and we subtract a pre-flare spectrum to obtain the flare-only flux.
 The pre-flare spectrum was calculated over
an average of 11 blue spectra before HST-1, and the scaling factor was
determined using increments of 0.005 relative to this pre-flare spectrum (see K16).  Because
the exposure times for the blue and red spectra differed
significantly, we averaged (four) red spectra to the times of each blue exposure;  the scaling obtained from the averaged red spectra was applied to each blue spectrum.
We verified that a continuum flux ratio ($F_{6010}/F_{6800}$) change occurs in the
spectra between 00:46:11 and 00:46:41 (corresponding to the peak of HST-1) before
applying the scale factor and subtracting the pre-flare.   We also
find that the equivalent width of H$\alpha$ closely follows its line-integrated flare-only flux.
These checks imply that the scaling algorithm is robust.
We also checked that the red continuum flare-only flux was greater than zero during
the flare and negligible before the pre-flare spectrum.  In Section \ref{sec:bj}, we show that the broadband increase from the red spectra during the peak of HST-1 is consistent with the data
in the $\sim V+R$ photometric band from the Helmos Observatory (Section \ref{sec:aristarchos}).

\clearpage
\subsection{Low-Resolution Spectroscopy from Keck 10m / LRIS} \label{sec:keckdata}
We obtained low-resolution spectroscopy from 3120 \AA\ to 5390 \AA\ with the 10m Keck/LRIS for one half night.  Inspired by the \citet{Herczeg2008} spectra of T Tauri spectra with LRIS, we aimed to measure the Balmer continuum shape down to the atmospheric cutoff during a flare.  Combined with the HST/COS data, the slope from 3100-3600 \AA\ would facilitate constraining the peak of the continuum (we do this in Section \ref{sec:combined}).
Three 30 second flats were obtained with a deuterium lamp, which is a bright ultraviolet continuum source, for the bluemost 300 pixels ($\lambda < 3645$ \AA).  Halogen flats were used for the redder pixels.  A master flat was made by combining these two flats.

We employed the atmospheric dispersion corrector (ADC) optimized for $3200-7000$ \AA, and we oriented the 1.5\arcsec\ slit at the parallactic angle (perpendicular to the horizon).  We binned 2x2 with low gain and fast readout 
resulting in 37~s between consecutive exposures with 45~s integration times.  The seeing was 0.73 \arcsec, and the conditions were clear.  The wavelength calibration was performed using a HgNeArZnCd lamp set, with Hg, Zn, and Cd providing the most lines in the NUV.  We achieved 0.3 \AA\ residuals with a cubic spline (order$=2$) fit to 15 arc features.  With the aid of the Keck/LRIS tool ARCPLOTS, the bluest feature that could be reliably identified is Cd 3261 \AA.  At bluer wavelengths between 3120 \AA\ and 3260 \AA, we relied on an extrapolation of the (low order, cubic spline) wavelength solution.  We obtained blue spectra with the 400/3400 grism, the 1.5\arcsec slit, the dichroic 560, and a dispersion of 2.1 \AA\ pixel$^{-1}$.  The spectral resolution, measured from the quiescent H$\delta$ emission line of GJ 1243, is $R\sim440$. 

We obtained five exposures of the spectrophotometric standard star BD$+$28 4211 at three values of airmass from 1.02 to 1.54.  The seeing was 
good enough to place the slit on the standard sdO star without contamination from the nearby fainter, redder star.  Using the calibrations from \citet{Oke1990} and the standard airmass extinction curve for Mauna Kea, we applied a second order extinction correction and flux calibrated the spectra. 

\begin{figure}
\begin{center}
\includegraphics[scale=0.35]{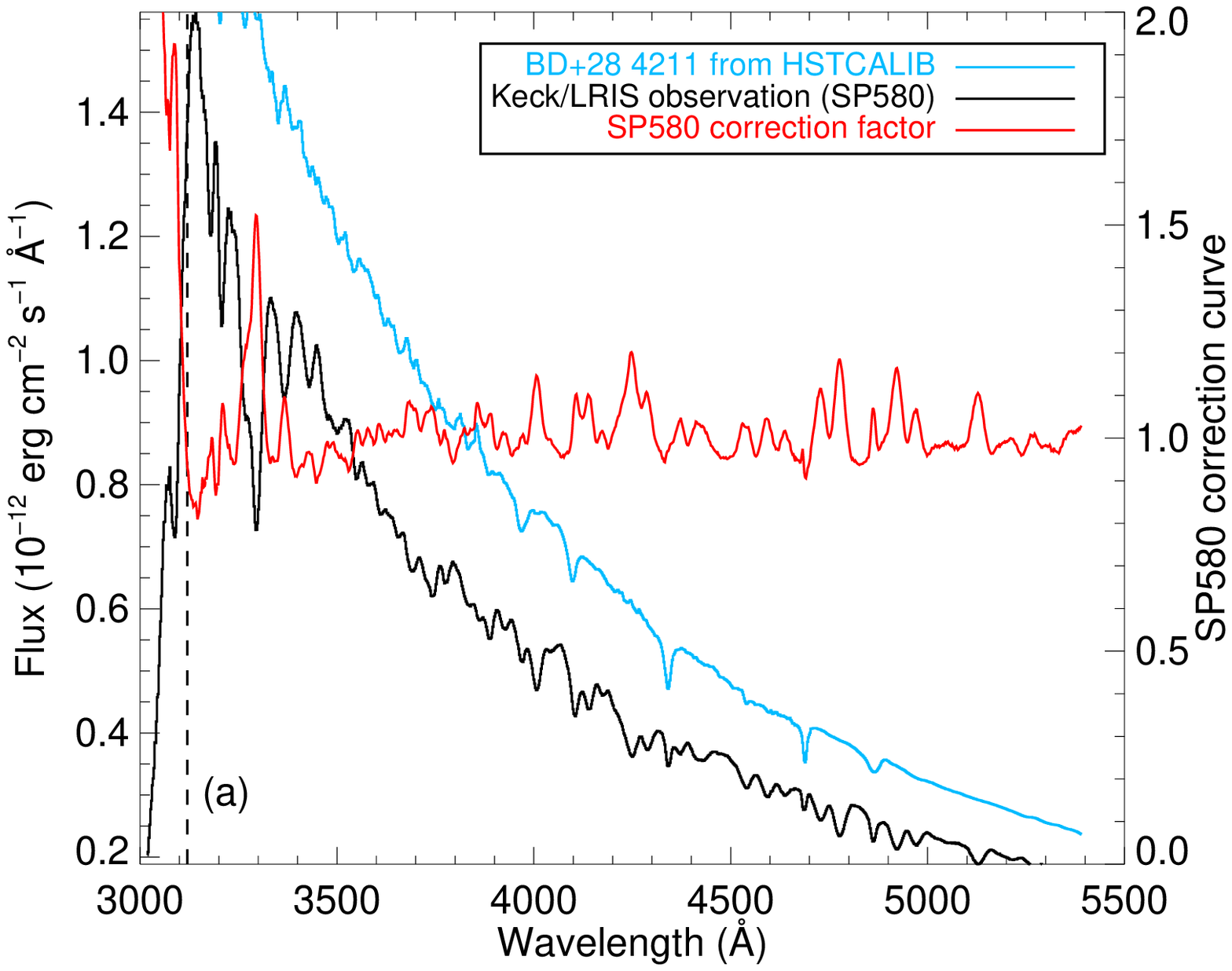}
\includegraphics[scale=0.35]{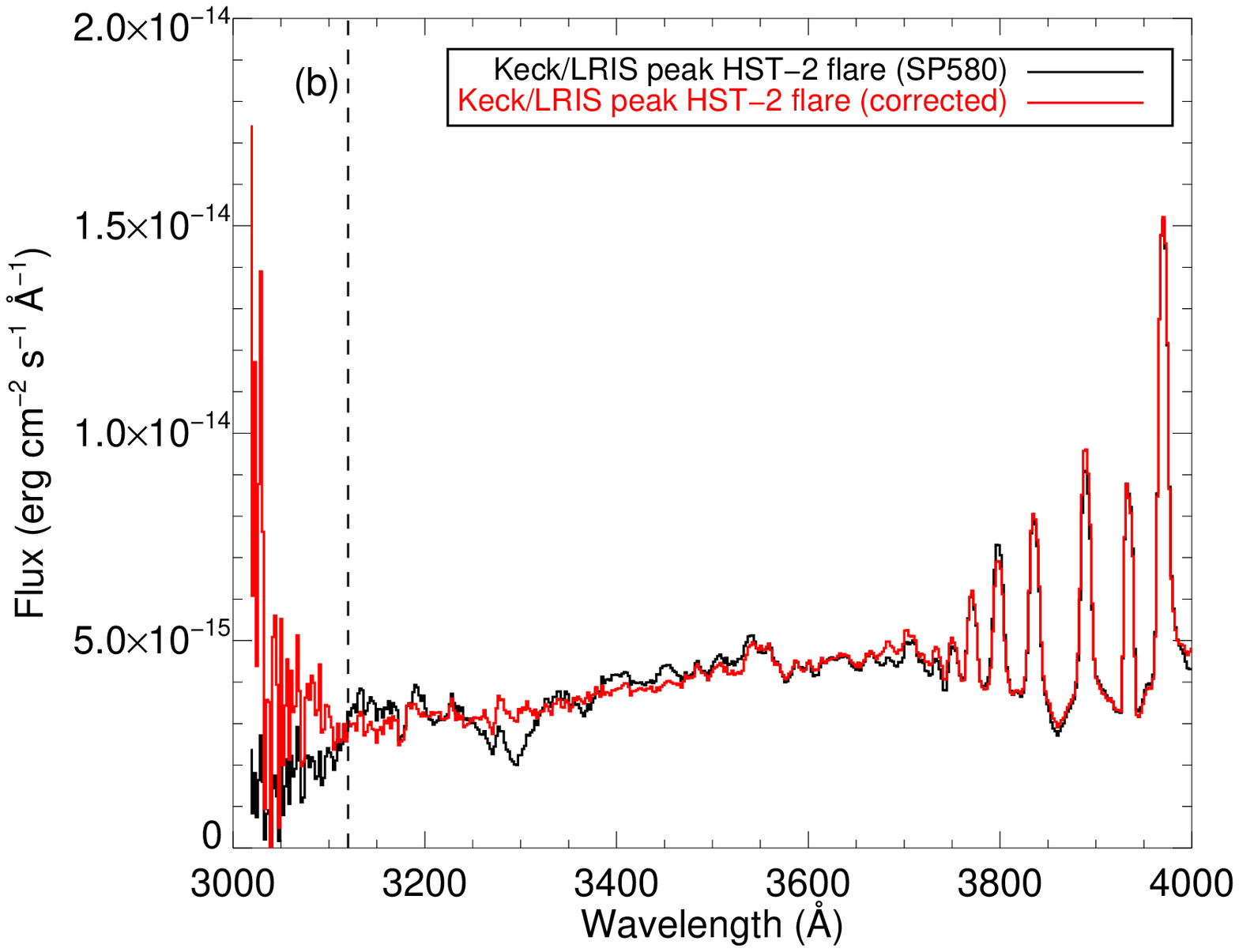}
\caption{\textbf{(a)}  Keck/LRIS observation of the spectrophotometric standard star BD$+$28 4211 taken at similar airmass and time as the HST-2 flare during the GJ 1243 observations.  The SP580 correction curve shows the values that are multiplied by the spectra of GJ 1243 to correct for the transmission of the SP580 filter.   \textbf{(b)}  The correction curve from panel (a) is multiplied by the HST-2 flare spectrum of GJ 1243 to obtain the corrected spectrum (\emph{red}).  The correction eliminates the fluctuations but does not affect the broad-wavelength shape of the continuum at $\lambda<3600$ \AA.   Vertical dashed lines indicate the bluemost wavelength (3120 \AA) that is calibrated using the spectrophotometric standard star flux.}
\label{fig:calib}
\end{center}
\end{figure}

At 5:22:40 UT (before all standard star and most GJ 1243 observations), the SP580 short-pass filter was inserted to block contamination in the blue from the red light of the star.  An observation of one of the BD$+$28 4211 exposures through the SP580 filter is shown in Figure \ref{fig:calib}(a).  The observation shows flux variations where there should be a continuum from the hot sdO star.  We found out that the variations are due to (undocumented) transmission curve wiggle-like variations in the SP580 filter, which was not used for the flat fields taken for this night and thus not able to be removed at early stages in the data reduction.  We scale the median of the CALSPEC observation of BD$+$28 4211 to the median of the Keck/LRIS observation and calculate a filter correction curve.  We apply this correction to all Keck/LRIS observations of GJ 1243.   Example flare-only spectra before and after this correction are shown in Figure \ref{fig:calib}(b).

The Keck Observatory obtained deuterium flats with the SP580 on 2015-02-13 for us but with a different CCD binning than our observations.  Additionally, the spectra became double-peaked due to focus problems during the main flare event.  Due to the problems with the observations on 2014-09-01, we caution that the Keck calibration potentially exhibits some intractable, but relatively minor, inaccuracies.  However, in Section \ref{sec:keck_lris}, we show the absolute flux calibration agrees well with the APO 3.5m spectra for overlapping wavelength coverage.  As these Keck spectra are among the very few \citep{Fuhrmeister2008, Fuhrmeister2011} flare observations down to the atmospheric limit, they are used for our analysis (Section \ref{sec:combined}) to demonstrate the value of future spectral observations at wavelengths at the atmospheric cutoff near 3200 \AA.  

The three spectra before the flare were averaged and scaled by the quiescent $B$-band magnitude of GJ 1243.  At each time during the flare, the flare-only spectra were calculated using the algorithm in \citet{Kowalski2013} for spectra with only blue-wavelength coverage. We analyze the Keck spectra S\#116 and \#117 which coincidentally cover the same times as S\#152-153 from APO.  The value of C3615\prim\ from this Keck flare spectrum is in remarkable agreement with the C3615\prim\ over S\#152 and S\#153 from APO (Section \ref{sec:keck_lris}), suggesting that the absolute flare-only flux levels are robust.

\subsection{$U$-band Photometry from the McDonald 2.1m} \label{sec:uband}

$U$-band data were obtained with the Otto Struve 2.1-m Telescope and the 
2-channel P45J photometer at the McDonald Observatory. Observations were 
obtained using a Johnson $U$ filter combined with a copper sulphate filter to 
eliminate the red leak present otherwise. This instrument provided a 
continuous sequence of photometric measurements without the time gaps  
necessary for CCD detector readouts. Integration times of 5 seconds were used.
One photometer was used to observe GJ 1243 through a 14.5\arcsec\ aperture, 
while a brighter, nearby, comparison star was observed with the second 
photometer through a smaller ($\sim$10\arcsec) aperture. This instrumental 
setup provided reliable differential photometry, even through light cloud. 
The main observational difficulty resulted from inaccurate telescope tracking 
at large hour angles but the resulting effects were mitigated by monitoring 
the comparison star signal. As the stars drifted, the comparison star 
reached the edge of its aperture first and GJ1243 was then recentered 
in its aperture. Recentering was performed just before the start of each 
HST orbit as a matter of routine. The sky signal was monitored frequently; 
typically at the start and end of each HST Earth occultation.

On 2014 Sep 01 data were collected between 02:22 and 08:24 UT with a 
cloud-free sky and several flares were recorded in the $U$-band light 
curve. The largest of these flares corresponded to the HST-2 flare 
observed by COS.

\subsection{$U$-band Photometry from the Isaac Newton Telescope} \label{sec:intdata}
Str\"omgren u-band (hereafter, denoted as $U$) data were obtained at La Palma with the 2.54-m Isaac Newton Telescope and the Wide Field Camera.
Exposure times were 20 seconds with 5.5 second for readout between exposures.  A binning of 2x2 was used.  The flat fields were obtained with a different
windowing than the observations.  Without flat fielding, we found that no larger than 5\% variations occurred on short (10 minute) timescales, and no larger than 10\% 
variations occurred on long (hour) timescales in the relative photometry of two comparison stars. The relative photometry on shorter timescales during quiescent times of GJ 1243 (just before 
the HST-1 flare) varies by only $\sim2.5$\%.  Thus we are confident that flares are robustly detected and characterized in these data.   The weather was mostly clear with high and decreasing humidity throughout the night.  
We used standard IRAF procedures to bias-subtract and perform aperture photometry with an aperture radius of 5 pixels, which is a factor of 1.3 larger than the FWHM of the target PSF at the time of the flare.  
	
\subsection{Photometry from the Helmos Observatory/Aristarchos 2.3-m} \label{sec:aristarchos}
GJ 1243 was observed on the night of 2014 Aug 31 with the RISE2 
instrument \citep{rise2} on the 2.3-m Aristarchos telescope at Helmos Observatory in 
Greece. The 1024x1024 E2V CCD47-20 back-illuminated CCD has pixels
that are 13 microns in size providing a pixel scale of 0.51\arcsec\ and a
field-of-view of 8.7\arcmin\ x 8.7\arcmin\ \citep{rise2}. The
exposure time was set to 5 s while the $V+R$ filter was used.  
Aperture photometry was extracted from GJ 1243 and the
4 brightest comparison stars available in the field, using a 7 pixel
or 3.6\arcsec\ radius with the IRAF apphot package and a 5 pixel wide
sky annulus 10 pixels from the source. 

\subsection{Additional Photometry and Spectroscopy} \label{sec:others}
We obtained three additional datasets that are not discussed in the analysis in this paper but are available upon request to the first author:
 $BVRI$ photometry from 2014-Sep-01 5:24 - 7:59 UT with the Prime Focus Corrector on the 0.8-m at the McDonald Observatory (exposure times were 30~s, 6~s, 2~s, 1~s
for $BVRI$, respectively);   low-resolution optical spectroscopy with the VIRUS-P spectrograph on the Harlan J. Smith 2.7-m telescope from 2014-Sep-01 2:11 - 9:35 UT at the McDonald Observatory (exposure times were 100~s); 
SDSS $gri$ photometry from 2014-Sep-01 2:21 - 10:30 UT with Flarecam on the  ARCSAT 0.5-m telescope at the Apache Point Observatory (exposure times were 30~s, 15~s, and 15~s for $gri$, respectively).

\section{Flare Light Curve Analysis} \label{sec:photanalysis}
A large fraction of the durations of the HST-1 and HST-2 flares were simultaneously observed from the ground-based telescopes and the HST.  
We use the high-cadence $U$-band light curves of the flares to characterize them and place them in context of previously observed dMe flares.  
To calculate flare energies in the $U$-band, we estimate the quiescent $U$-band flux ($2.9\times10^{-15}$ \cgs\ ), using the $B$-band mag (14.47) and the $V$-band mag (12.83) of GJ 1243 from \citet{Reid2004} and the zeropoint flux density values from \citet{Bessel2013}.  We follow \citet{Hawley2014} and assume the $U-B=0.93$ of YZ CMi since there is no published value of the $U$-band magnitude for GJ 1243.   From our quiescent spectra (Section \ref{sec:keck_lris}), we estimate a $U$-magnitude of $\sim15.5$.  However, a precise conversion from spectrophotometry to $U$-band photometry is generally not possible due to uncertainties in the blue response of the telescope, instrument, and atmosphere that factor into any system's total $U$-band response function.  The spectrophotometric $U$-band magnitude of GJ 1243 results in $\sim10$\% lower inferred flare energies, which is not critical for our analysis.   Using the $U-B$ color of YZ CMi facilitates a direct comparison to the flare energies in \citep{Hawley2014}, and we note that the NUV is also rather variable in quiescence (see Section \ref{sec:combined}). 

We performed our analysis using a distance of 12.05 pc \citep{Harrington1980}. Subsequently, Gaia DR2 \citep{Gaia, GaiaDR2} published a parallax distance of 11.9787 $\pm$ 0.0052. This 0.5\% difference in distance (1\% difference in flare energy) is negligible for our analysis.
From the broadband $U$-band and NUV light curves, we calculate flare energies, flare peak flux enhancements ($I_f+1$), the FWHM of the light curves ($t_{1/2}$), and the impulsiveness indices ($I_f/t_{1/2}$) from K13.  Some properties of the NUV light curves are given in Table \ref{table:hstdata}.

\clearpage
\begin{itemize}
\item HST-1: 

Figure \ref{fig:lcfigs}(a) shows the NUV light curve for the HST-1 flare at 5~s intervals. 
This flare exhibits two to three fast-rise events in the impulsive phase, which produces a maximum NUV flux enhancement of $I_f+1 = 9.5$.  The 60~s integration times of the WHT spectra and the NUV light curves extracted from these times are shown on this figure.  We focus our WHT spectral analysis on S\#163, which corresponds to the times covering the peak, initial fast decay, and initial gradual decay phases.

The $U$-band energy of this event is $4.4\times10^{31}$ erg, with 35\% of the energy radiated in the impulsive phase.  The total flare duration (including several smaller $U$-band events in the gradual decay) is $\sim1800$~s.  The peak $U$-band flux enhancement is $I_f+1\sim 4.3$ ($\Delta U \sim -1.6$ mag, at $t_{\rm{exp}}=20$~s) and the value of $t_{1/2}=135$~s, giving a $U$-band impulsiveness index of $\sim1.5$.  

\begin{figure}
\includegraphics[scale=0.45]{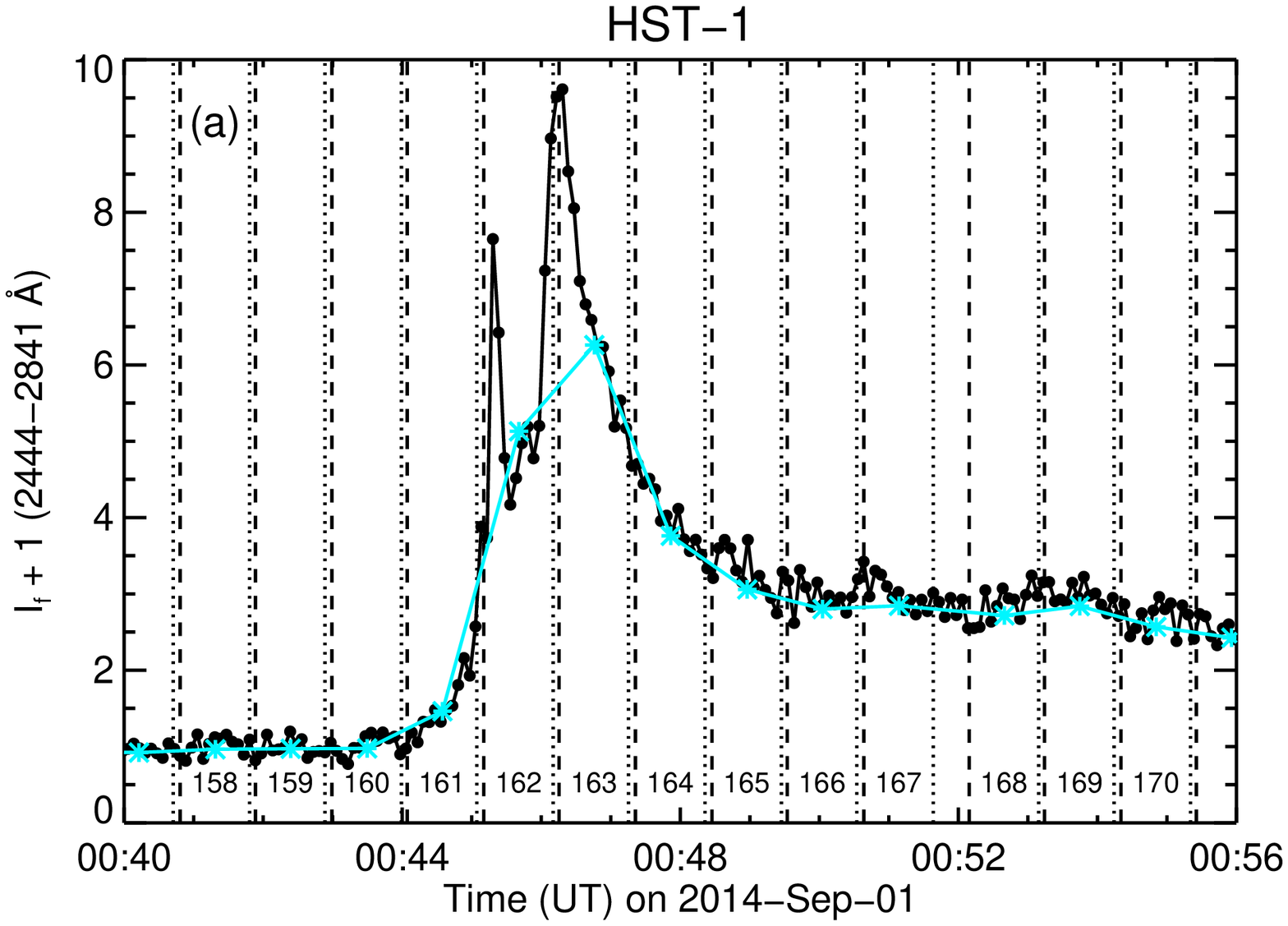}
\includegraphics[scale=0.45]{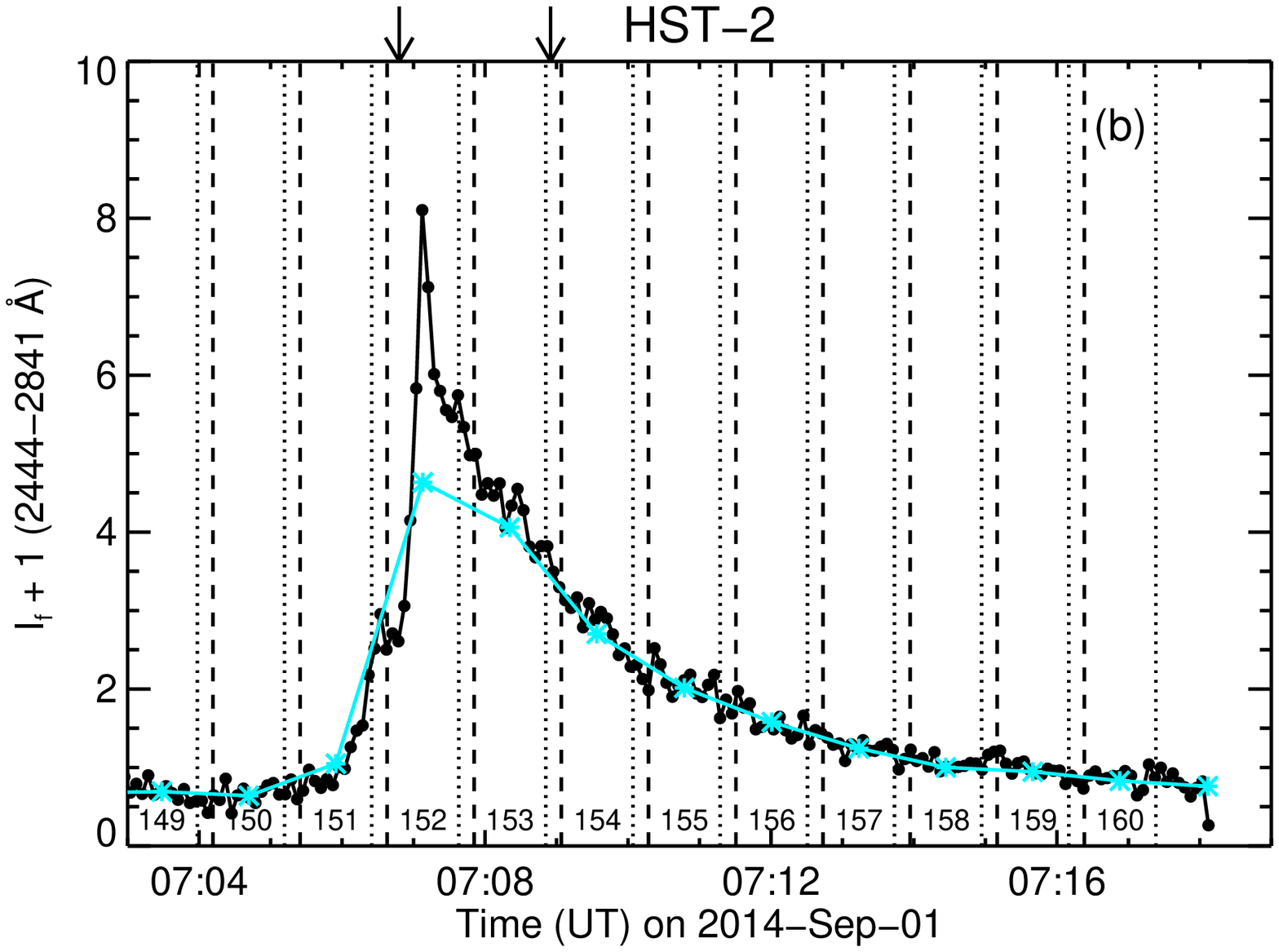}
\caption{\textbf{(a)} Wavelength-integrated  (2444-2841 \AA) NUV light curve of the HST-1 flare event over 16 minutes of HST observations binned by 5~s.  At 00:56 UT, GJ 1243 was occulted by the Earth.  The spectral integration times of the ISIS spectra from the WHT 4.2-m are indicated as vertical lines (exposure start by a vertical dashed lines, exposure end by vertical dotted lines). The sequential numbering of the spectra along the bottom (S\#) is used throughout the text.  \textbf{(b)}  Wavelength-integrated NUV light curve of the HST-2 flare event at 5~s time-binning over 16 minutes of COS observations.  The spectral integration times of the DIS spectra from the ARC 3.5-m are indicated as vertical lines (exposure start by a vertical dashed lines, end by vertical dotted lines). The sequential numbering of the spectra along the top (S\#) is used throughout the text.  Arrows indicate the times over which Keck/LRIS spectra S\#116 and S\#117 are averaged and analyzed in Section \ref{sec:combined}.  Note that the pre-flare flux is 70\% of the pre-flare flux
in panel (a), but both flares are normalized to the same quiescent wavelength-integrated NUV flux value ($3.54\times10^{-13}$ erg cm$^{-2}$ s$^{-1}$) obtained just before HST-1 (the varying level of quiescent flux can be seen in Figure \ref{fig:hstlc}).  Since the two light curves are normalized to the same quantity, the specific luminosity the flares can be directly compared.  Both panels show the wavelength-integrated ($\lambda=2510-2841$ \AA) NUV light curves as \emph{cyan asterisks} extracted at the exposure times of the respective ground-based spectra.   }
\label{fig:lcfigs}
\end{figure}

\item HST-2: 

Figure \ref{fig:lcfigs}(b) shows the NUV light curve for the HST-2 flare at 5~s intervals, to be compared directly to HST-1 in panel (a).  This flare exhibits a peak NUV flux enhancement of $I_f+1 = 8$. This flare also has several fast events in the impulsive phase, though these fast events have different relative amplitudes compared to those in the impulsive phase of HST-1.  The 60~s integration times of the APO spectra and the NUV light curves extracted from these times are shown on this figure.  We focus our APO spectral analysis on the average of S\#152 and S\#153, which correspond to the fast rise, peak, fast decay, and initial gradual decay phases. 

The 5~s cadence $U$-band light curve covering HST-2 is shown in Figure \ref{fig:uband}.  Note that eleven other flares with amplitudes $I_f+1<2$ occur over the 5.73 hours of these $U$-band observations.  As seen in the NUV light curve (Figure \ref{fig:lcfigs}(b)), the HST-2 event exhibits two $\sim20$~s episodes of fast rise over a total impulsive phase duration of 80~s, with a peak flux enhancement of $I_f+1=3.7$ ($\Delta U_{\rm{peak}} = -1.4$ mag) that is followed by a short fast decay phase and a much longer gradual decay phase that 
begins at 80\% of the flare maximum.  The total flare duration is 600 s, but HST went behind the Earth before the flare was completely over, as for HST-1. The $U$-band equivalent duration \citep{Gershberg1972} of 450s gives a $U$-band energy of 1.6$\times10^{31}$ erg, placing it very similar in energy to the 
$U$-band flare on GJ 1243 reported in \citet{Hawley2014} and in the middle of the energy distribution of flares on GJ 1243 \citep{Hawley2014}.  The peak $U$-band luminosity is nearly $10^{29}$ erg s$^{-1}$.  In contrast to HST-1, about 1/5 of the total energy is radiated in the impulsive phase.
At 5~s cadence, the $t_{1/2}$ value is 120~s, which gives a $U$-band impulsiveness index of 1.4 (or 0.9 when the 5~s U-band light curve is binned to 20~s), which is similarly small like the impulsiveness of HST-1.

\end{itemize}

\begin{deluxetable}{lcccccccccc}
\rotate
\tabletypesize{\scriptsize}
\tablewidth{0pt}
\tablecaption{NUV Flare Peak Properties}
\tablehead{
\colhead{Flare} &
\colhead{S\#} &
\colhead{$I_f+1$ at peak*} &
\colhead{$t_{1/2}$ (s)*} &
\colhead{$t_{\rm{fast}}$ (s)*} &
\colhead{$t_{\rm{imp}}$ (s)*} &
\colhead{Fraction energy in continuum} &
\colhead{Fe II / C2650\prim} & 
\colhead{Mg II / C2650\prim} &
\colhead{Fe II / Mg II} &
\colhead{C2650\prim\ / C2820\prim\ (err)} }
\startdata 
HST-1 & 163 (WHT) &  9.5 &  65 & $<5-15$ & 95 & 0.62 & 20 & 46 &  0.43 & 0.94 (0.11) \\ 
HST-2 & 152-153 (APO) &  8 & $60-75$ & 15 & 60 & 0.65 & 18 & 39 & 0.46  & 0.88 (0.08) \\
\enddata 
\tablecomments{*From high-cadence ($t_{\rm{exp}}=5$~s) light curves (Figures \ref{fig:lcfigs}(a)-(b)).  $t_{\rm{fast}}$ is the duration range for the times of fast rise in the light curves.  $t_{\rm{imp}}$ is the duration of the impulsive phase including all bursts in the light curves.  Fe II refers to the line-integrated flux of the four emission lines Fe II $\lambda2599.15$, $\lambda2600.17$, $\lambda2631.83$, and $\lambda2632.11$.  Mg II refers to the  line-integrated flux of Mg II $\lambda2791.6$,  $\lambda2796.35$ ($k$), $\lambda2798.8$, and $\lambda2803.5$ ($h$).  Prime symbols indicate flare-only specific flux values in continuum regions.  }
\end{deluxetable}\label{table:hstdata}

According to the high-cadence $U$-band ($t_{\rm{exp}}=5-20$~s) impulsiveness index categorization from K13, the HST-1 and HST-2 flares are hybrid flare (HF) events; they fall between HF1 and HF2 in the sample of K13.  HST-1 and HST-2 have similar peak flux enhancements and two to three episodes of fast rise in their respective impulsive phases.  However, HST-1 is several times more energetic, exhibits a slower rate of gradual decay, and is three times longer total duration.

At a few $\sim10^{31}$ erg in the $U$-band, the HST-1 and HST-2 events likely have bolometric white-light flare energies that are comparable to (or greater than) the total white-light energy produced in the largest X-class flares observed in the Sun \citep{Neidig1994, Woods2004, Kretzschmar2011, Osten2015}.  
\citet{Kosovichev2001} has shown that the famous 2000 July 14 Bastille Day solar flare has a double-peaked optical light curve, and each peak is attributed to a spatially distinct, but neighboring and magnetically connected, set of two ribbons.  \citet{Qiu2010} has shown that this flare consists of two phases of magnetic field unshearing, rapid spreading of ribbons, and enhanced reconnection rate.  Perhaps the two flares on GJ 1243 each consisted of two spatially adjacent, two-ribbon events that produced the multiple episodes of fast rise in the impulsive phases, as in the Bastille Day solar flare event.   A more detailed comparison to solar flares will be presented in a future work.

\begin{figure}
\includegraphics[scale=0.3]{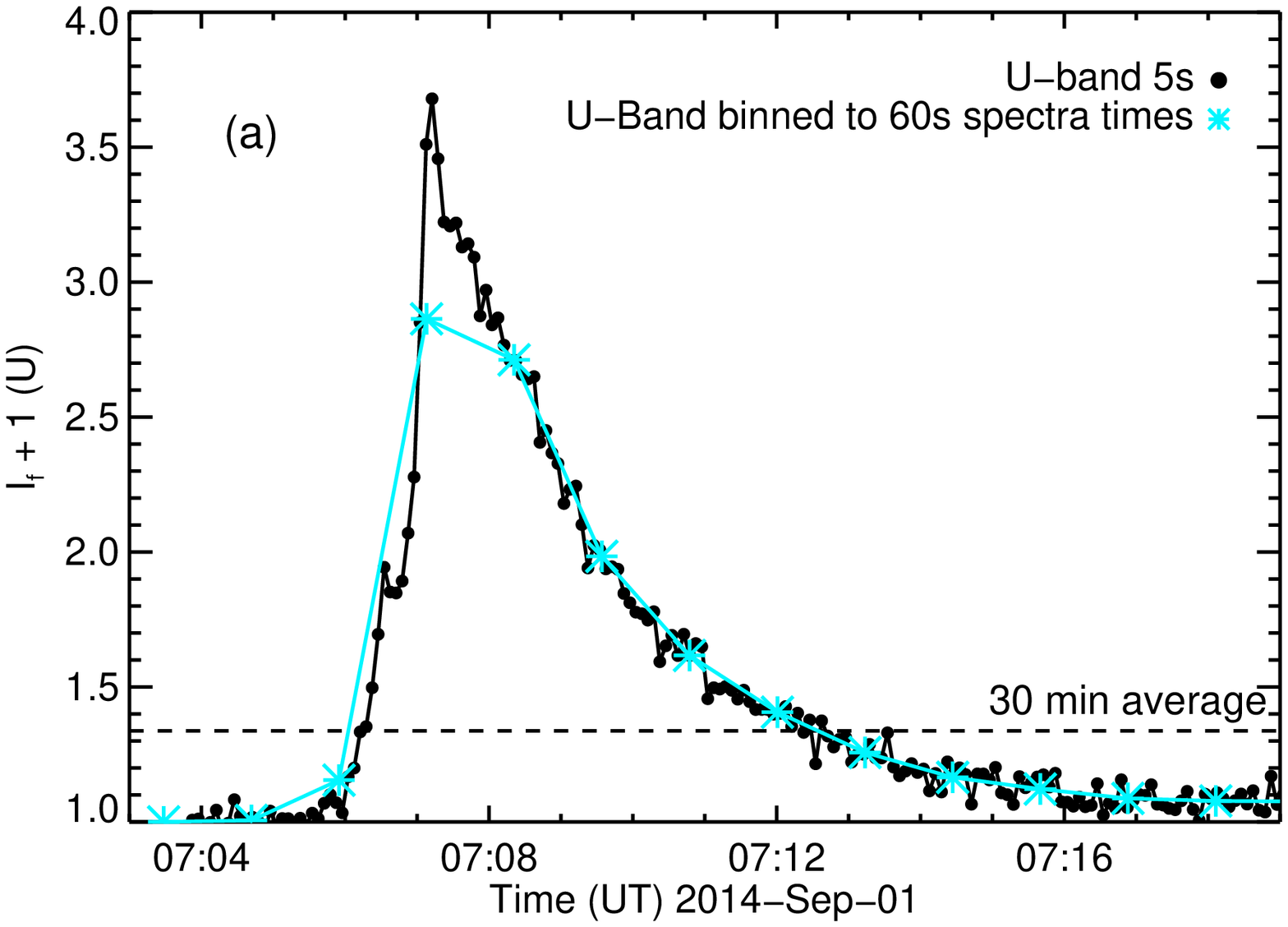}
\includegraphics[scale=0.3]{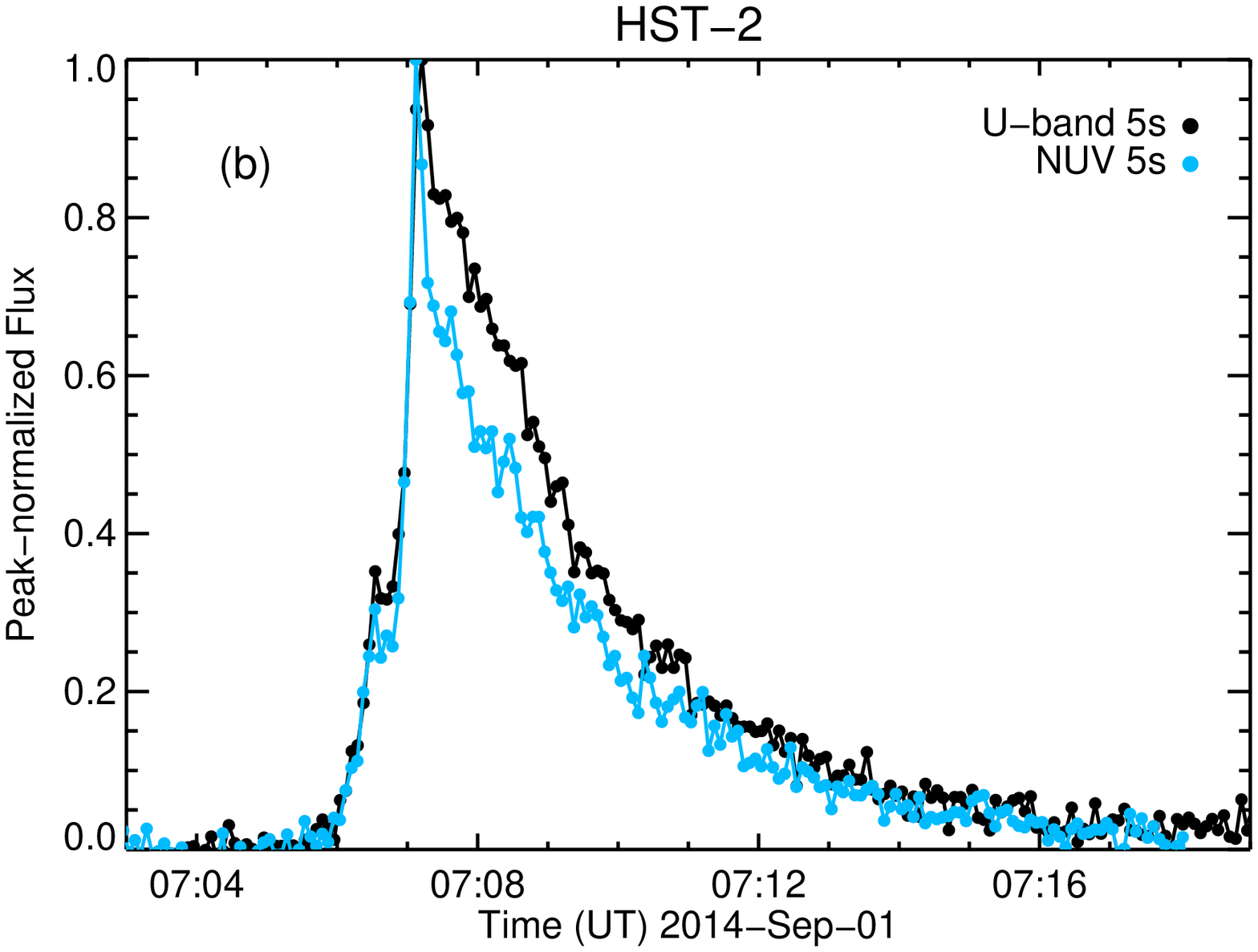}
\includegraphics[scale=0.3]{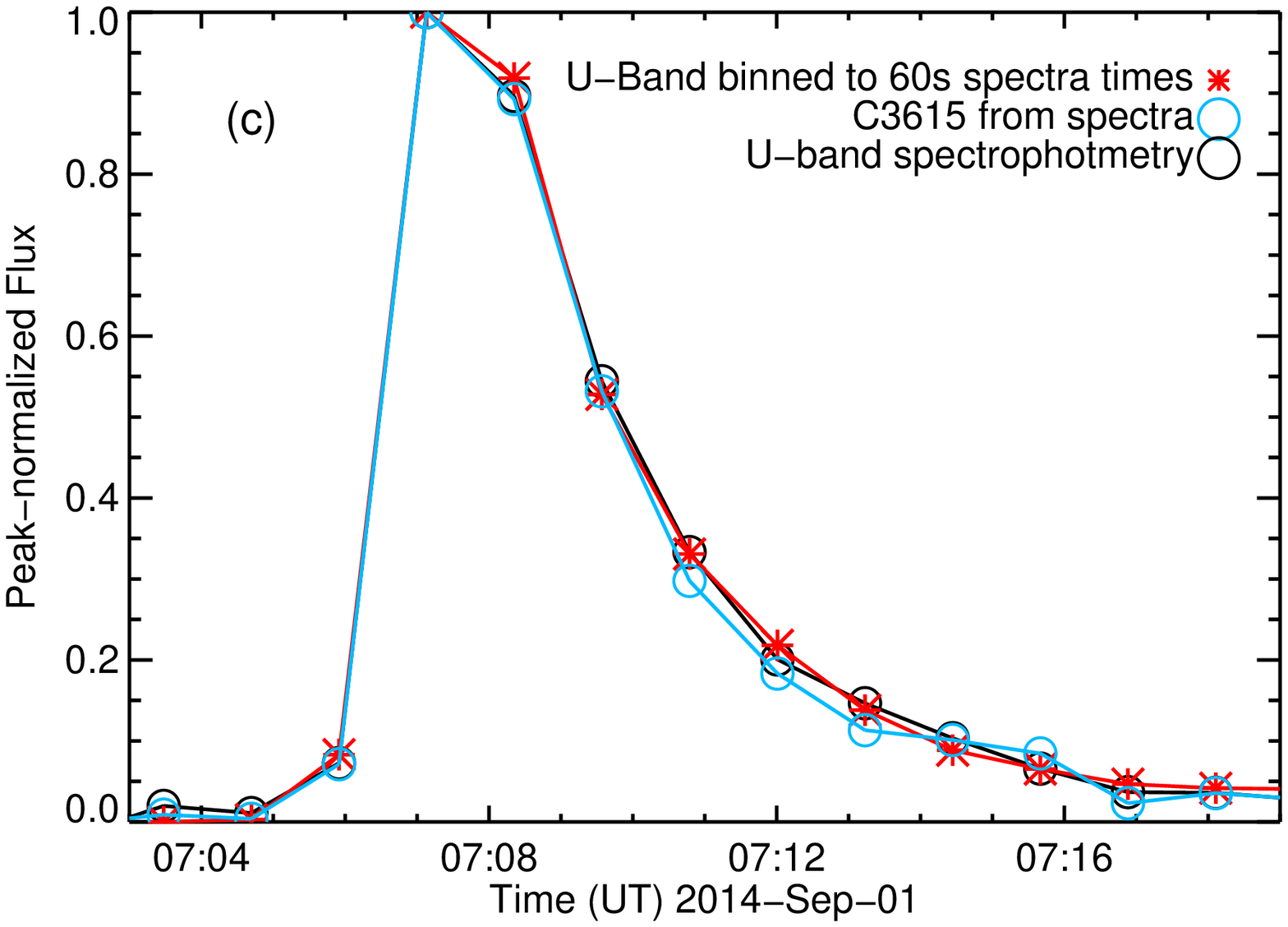}
\caption{ \textbf{(a)} The $U$-band light curve of HST-2 at its original 5~s cadence, at a 60~s cadence, and a 30-minute average.   \textbf{(b)} Comparison of HST-2 in the $U$-band and NUV at the same high-time ($t_{\rm{exp}}=5$~s) cadence; the light curve of NUV from Figure \ref{fig:lcfigs}(b) is reproduced in \emph{blue circles}.  The peak-normalized flux values have the pre-flare
levels subtracted before dividing by peak values.  The NUV is slightly faster to reach peak and to reach the gradual decay phase than the $U$-band; thus the NUV has shorter values of $t_{1/2}$ (see Table \ref{table:hstdata}). \textbf{(c)} The peak-normalized $U$-band photometry and the $U$-band spectrophotometry (from the 3.5-m/DIS spectra) are in satisfactory agreement when binned to 60~s exposure times.  The value of C3615\prim\ from the 3.5-m/DIS spectra also follows the $U$-band.}
\label{fig:uband}
\end{figure}

\subsection{The $U$-band \emph{vs.} NUV:  similarities and differences} \label{sec:nuvdiff}
The light curves of HST-1 and HST-2 in Figure \ref{fig:lcfigs} demonstrate the importance of high-time resolution with $t_{\rm{exp}}=5$~s or better during flares. 
Figure \ref{fig:uband} shows several more light curves of HST-2.  In Figure \ref{fig:uband}(a), the $U$-band photometry from the Otto Struve Telescope \label{sec:uband} is binned to the ground-based spectral integration times of 60~s (\emph{teal/cyan asterisks}), which
follows the 60s-binned light curve of the wavelength-integrated flux in Figure \ref{fig:lcfigs}(b).  Clearly, a 1-minute average obscures light curve detail and confuses the identification of the distinct phases of the flare:  at relatively low cadence, two points appear as the impulsive phase while the fast decay phase corresponds to
the gradual decay phase in the higher cadence data.  In \emph{Kepler} 30-minute cadence data \citep[e.g.][]{Walkowicz2011, Shibayama2013, Davenport2016, Yang2017, VanD}, the flare would be a single point at the level indicated in the plot.  From the lower-cadence ($t_{\rm{exp}}=60$~s) light curves, the HST-1 and HST-2 events are classified as hybrid flare (HF) events
with impulsiveness indices of $0.6-0.8$ (K13).  

When we compare the high-cadence $U$-band data to the high-cadence NUV light curve of HST-2 (Figure \ref{fig:uband}(b)), we find that the NUV peaks slightly before and decays faster in the fast decay phase.  Although the $U$-band data of HST-1 is lower cadence ($t_{\rm{exp}} = 20$~s) and was not reduced with a flat field (Section \ref{sec:intdata}), the peak-normalized NUV light curve decays slightly faster than the $U$-band light curve, as in the HST-2 event.

  The difference in timescales between the two regimes of the NUV is surprising and should be investigated further with high-cadence observations and 
pursued with radiative-hydrodynamic flare modeling.  We speculate that possible sources of the differences could be 1) relatively more Balmer continuum radiation in the $U$-band, which also has a contribution from high order Balmer lines, and/or 2)  a moderately high and variable optical depth as a function of the wavelength in the Balmer continuum from $\lambda=2600$ \AA\ to the Balmer series limit at 3646 \AA.   
In Figure \ref{fig:uband}(c) we compare the 60s-binned $U$-band observations to two quantities obtained from the ARC 3.5-m/DIS spectra (Section \ref{sec:apodata}), which are discussed in the next section (Section \ref{sec:bj}).

\section{Balmer Jump Spectral Analysis} \label{sec:bj}
The interpretation of the continuum radiation in the NUV range critically depends on the properties of the Balmer jump and optical continuum radiation constrained by the ground-based telescopes.  
In this section, we analyze the Balmer jump spectra covering the peaks and initial decay phases of HST-1 (S\#163 from the WHT in Figure \ref{fig:lcfigs}(a)) and of HST-2 (an average of S\#152-153 from the ARC 3.5m in Figure \ref{fig:lcfigs}(b)).  
Spectral quantities described in K13 and K16 are calculated from these spectra and are given in Table \ref{table:data}.

\begin{figure}
\includegraphics[scale=0.50]{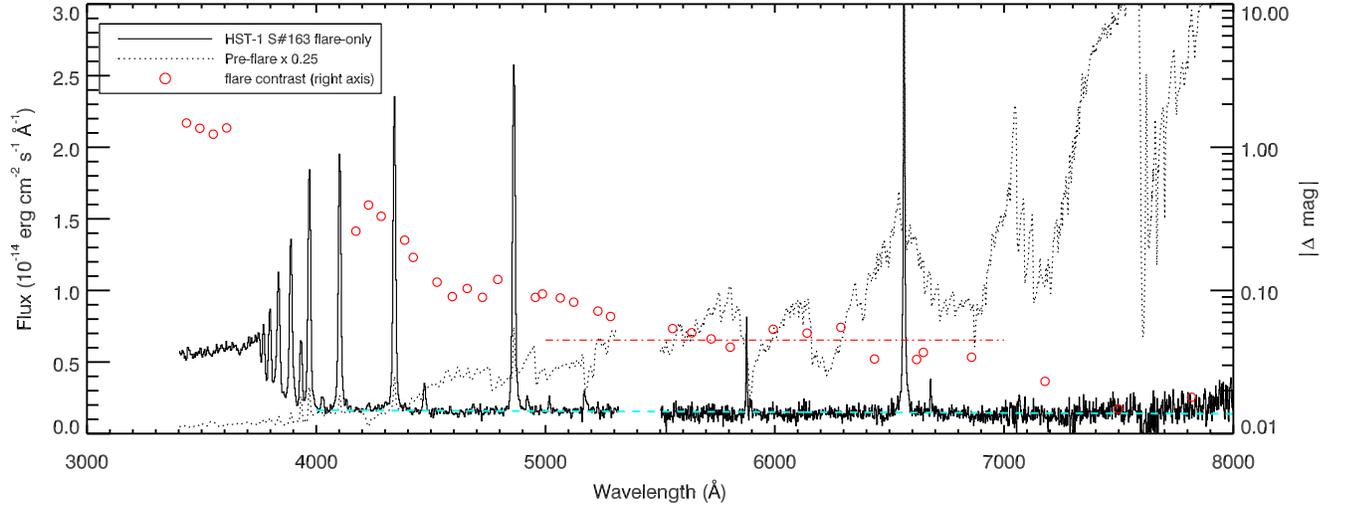}
\caption{HST-1 flare spectrum over the peak and initial decay phases at $\lambda=3400-8000$ \AA.  Four 10-second red spectra from the WHT are coadded for the time corresponding to the 60s exposure S\#163 in the blue from the WHT.  The \emph{red circles} indicate the magnitude changes (flare contrast; right axis) within continuum-only wavelength regions:  the dot-dashed red line shows the average magnitude change in the spectrum over the filter of the Aristarchos photometry (Figure \ref{fig:aristarchos}).   The flare-only continuum from $\lambda=4000 - 8000$ \AA\ is nearly flat and is fitted with a line (\emph{dashed cyan}).   The non-flaring spectrum scaled by 0.25 is shown as a \emph{dotted line}.     }
\label{fig:fullsed}
\end{figure}

\begin{figure}
\begin{center}
\includegraphics[scale=0.55]{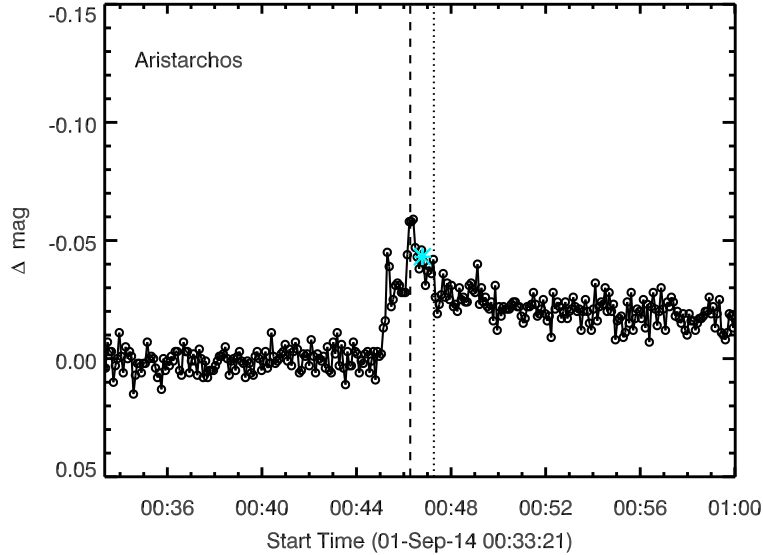}
\caption{ Light curve of GJ 1243 from Aristarchos in $\sim V+R$ band.  The \emph{cyan asterisk} shows the average photometry over the exposure time of S\#163 from the WHT.   Note, the photometry does not return to the pre-flare level for the time-range shown. }
\end{center}
\label{fig:aristarchos}
\end{figure}

The peak flare-only spectrum of HST-1 (S\#163; Figure \ref{fig:lcfigs}(a)) at $\lambda=3400-5300$ \AA\ in the $U$-band, the blue-optical, and the red optical ($\lambda=5500-8000$ \AA) are shown in Figure \ref{fig:fullsed}.  To our knowledge, Figure \ref{fig:fullsed} is the first robust characterization through the full optical wavelength regime near the peak of a moderate-amplitude, HF event on a dMe star.  Near the peak (S\#163), HST-1  exhibits a moderate Balmer jump ratio,  $\chi_{\rm{flare}}=$C3615\prim/C4170\prim\ of  $3.6\pm0.3$.  No significant spectral variation (except in total flux) occurs from S\#162 to S\#163.  
 On the right axis of Figure \ref{fig:fullsed} we show the contrast ($\Delta$mag) of the flare-only continuum windows.  From the flare spectrum, we estimate the 
magnitude change over the wavelengths corresponding to the bandpass of the Aristarchos photometry ($\sim V+R$ band), which is shown in Figure \ref{fig:aristarchos}.  The agreement of $-0.045$ mag is remarkable\footnote{We note that the H$\alpha$ line flux accounts for only $\sim13$\% of the $\lambda=5600-7000$ \AA\ flare-only flux, which is an estimated $\Delta\rm{mag} \sim -0.005$ in the Aristarchos bandpass.}.   With this verification of the flux calibration, we measure the blue-to-red flux ratio, C4170\prim/C6010\prim, to be 1.15$\pm0.24$ in the flare spectrum in Figure \ref{fig:fullsed}.  The optical continuum is nearly flat, and it is fit by an (unweighted) line with a nearly flat slope in continuum-only wavelength windows at $\lambda=4000-8000$ \AA\ in Figure \ref{fig:fullsed}.  The value of C4170\prim/C6010\prim\ calculated from the best-fit line is 1.08 and is consistent with the continuum flux ratio calculated from the data.  We also calculate the percentage of $4000-8000$ \AA\ energy under the line fit (\emph{light blue line}) to be 80\% of the flare-only energy, and the $\lambda=3420-5200$ \AA\ HB percentage\footnote{The HB quantity is defined in K13.  If we do not subtract a linear fit from the Balmer continuum wavelengths, the HB percentage is 60\%.} to be 48\%, which is similar to the impulsive/peak phase properties of other GF-type events in K13 (cf their Fig 21).

\begin{figure}
\begin{center}
\includegraphics[scale=0.70]{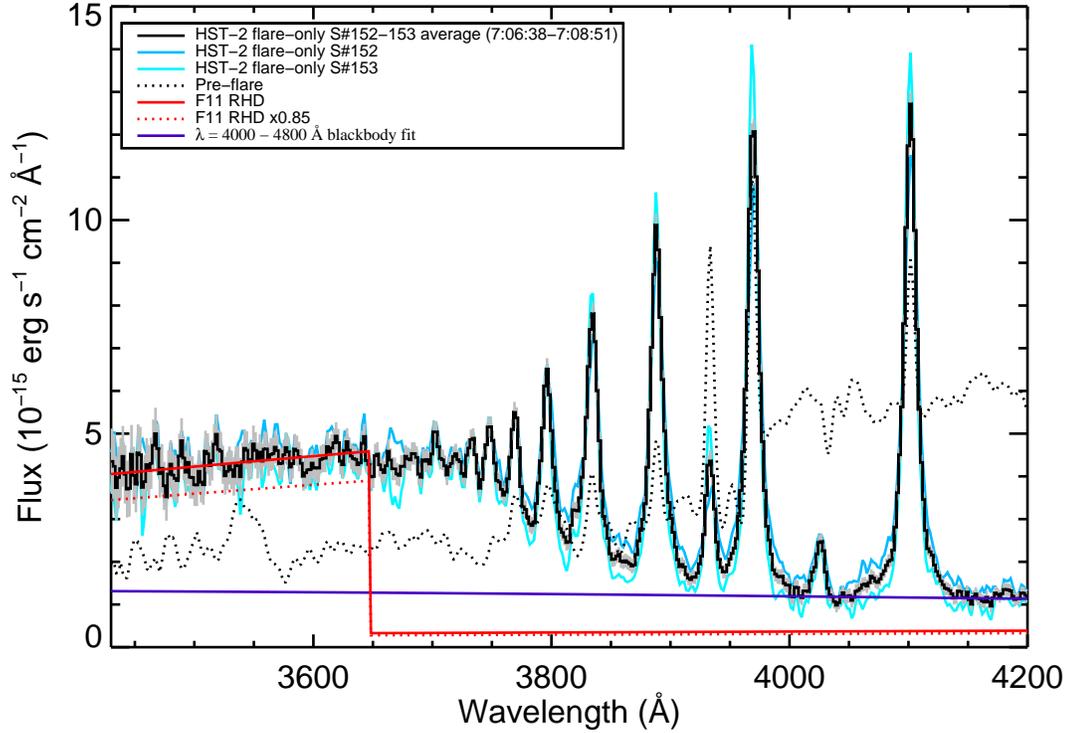}
\caption{Impulsive phase Balmer jump flare spectra of the HST-2 event, shown from $\lambda=3430-4200$ \AA.  A blackbody with $T_{\rm{BB}}=9000$ K is shown as the \emph{purple} curve, which is fit to the \emph{black} flare spectrum;  the F11 RHD model ($t=2.2$~s) is shown in \emph{red} scaled to the observation; the \emph{dotted red} line shows the RHD F11 model that is scaled by eye to the bottom troughs in the observed flux at $\lambda < 3600$ \AA. The \emph{dotted black} line is the pre-flare (background) spectrum of GJ 1243.   The calculated quantities from the average of S\#152 and S\#153 (\emph{black} with \emph{grey} error bars) spectrum are given in Table \ref{table:data}.   }
\end{center}
\label{fig:balmerjump}
\end{figure}

In Figure \ref{fig:balmerjump} we show the spectra S\#152, S\#153, and the average of S\#152-153 at $\lambda > 3430$ \AA\ from APO/DIS for HST-2, which showcases
the Balmer jump properties of this flare and a hot, $T_{\rm{BB}}\sim9000$ K, blackbody continuum that is fitted to the blue optical ($4000-4800$ \AA) wavelength regime of the average of S\#152 and S\#153.  The observed spectra S\#152 and S\#153 include the rise, peak,
and post-peak phases of the flare (Figure \ref{fig:lcfigs}(b)).  Though high-time resolution is important, we coadd S\#152 to S\#153 in order to attain a robust estimate of a color temperature at blue-optical wavelengths where the flare contrast against the molecular band pseudo-continuum in the preflare spectrum is low ($\le 20$\%). 
The Balmer jump ratio, $\chi_{\rm{flare}}$, is $3.8 \pm0.4$ in the coadded spectrum, which does not vary significantly between
the S\#152 (3.5$\pm0.4$) and S\#153 (4.2$\pm0.9$) spectra\footnote{Despite the issues with the calibration of the Keck/LRIS spectra, we calculate a similarly moderate Balmer jump ratio of $\sim3$ at the peak of HST-2.}, following the calculation of flare color errors in K13\footnote{If we use the error on the weighted mean (instead of the standard deviation of the flux within each continuum window as in K13) we obtain an uncertainty of $0.17$ and 0.24 on the Balmer jump ratios for S\#152 and S\#153, respectively, which suggests that there is significant intraflare variation with a larger Balmer jump ratio in the gradual decay phase.  This is qualitatively consistent with the evolution of the Balmer jump ratios for other dMe events in K13.  In Section \ref{sec:combined}, we calculate the flare color errors from these spectra as the error of the weighted means with a systematic flux calibration uncertainty of 5\% added in quadrature.  } 
   The Balmer lines are broad in this flare.  Moreover, they are bright relative to the continuum, exhibiting a value of H$\gamma$/C4170\prim\ of 160.  The last identifiable Balmer line is H14 $\lambda3722$, while He I 3705 and He I 4026 are in emission, and Ca II K is characteristically fainter during the flare than the nearby Balmer lines (e.g., H8).   In this event, the value of $\chi_{\rm{flare}}$, H$\gamma$/C4170\prim, the HB percentage, and BaC3615/C3615\prim\ are all very similar to the respective quantities in HST-1 in Figure \ref{fig:fullsed}.

The F11 RHD continuum model spectrum from \citet{Kowalski2015} is shown and scaled to match the observed flare-only flux at $\lambda=3500-3640$ \AA. 
The overall shape at $\lambda < 3646$ \AA\ is well-represented by this RHD continuum spectrum, which results from hydrogen recombination radiation formed over low continuum optical depth.
However, the observed Balmer jump ratio is not reproduced in the RHD model, which has a value of $\chi_{\rm{flare}}=11$.  The RHD model is scaled to the lowest possible values in the observed spectra \citep[qualitatively like the fitting method of][]{Milligan2014} and is also shown (\emph{dotted red line}).  The blending of the higher order Balmer lines and the dissolved level pseudo-continuum
at $\lambda=3646-4000$ \AA\ will be addressed in Paper II using the modeling techniques in \citet{Kowalski2017B}.

 The evolution of C3615\prim\ and the $U$-band that is synthesized from the spectra using the transmission curve in \citet{Bessel2013} are shown in Figure \ref{fig:uband}(c) compared to the 60s-binned $U$-band light curve.  They follow each other remarkably, which demonstrates the
 robustness of our calculation of the flare-only flux for moderate amplitude flares (Section \ref{sec:apodata}).  In this figure, the light curves are normalized to their peak values, and the spectrophotometric $U$-band peak  value of $I_f+1$ is
 within 10\% of the 2.1-m/$U$-band light curve peak flux enhancement ($I_f+1 \sim 3$; Figure \ref{fig:uband}(a)) binned to 60 s.
 HST-2  exhibits a flux enhancement of $\sim$3 in the NUV at C3615\prim, but there is only a flux enhancement of 1.2 in the blue continuum at C4170\prim.  This low flare contrast in the blue makes this flare is a good case to combine with the constraints from shorter wavelengths to test RHD model predictions and the extrapolation of a $T=9000$ K blackbody (Section \ref{sec:combined}).  

\section{Discussion I: The HST-1 and HST-2 flare events in the context of other \lowercase{d}M\lowercase{e} flares with Balmer jump spectra} \label{sec:discussion1}
\begin{figure}
\begin{center}
\includegraphics[scale=0.70]{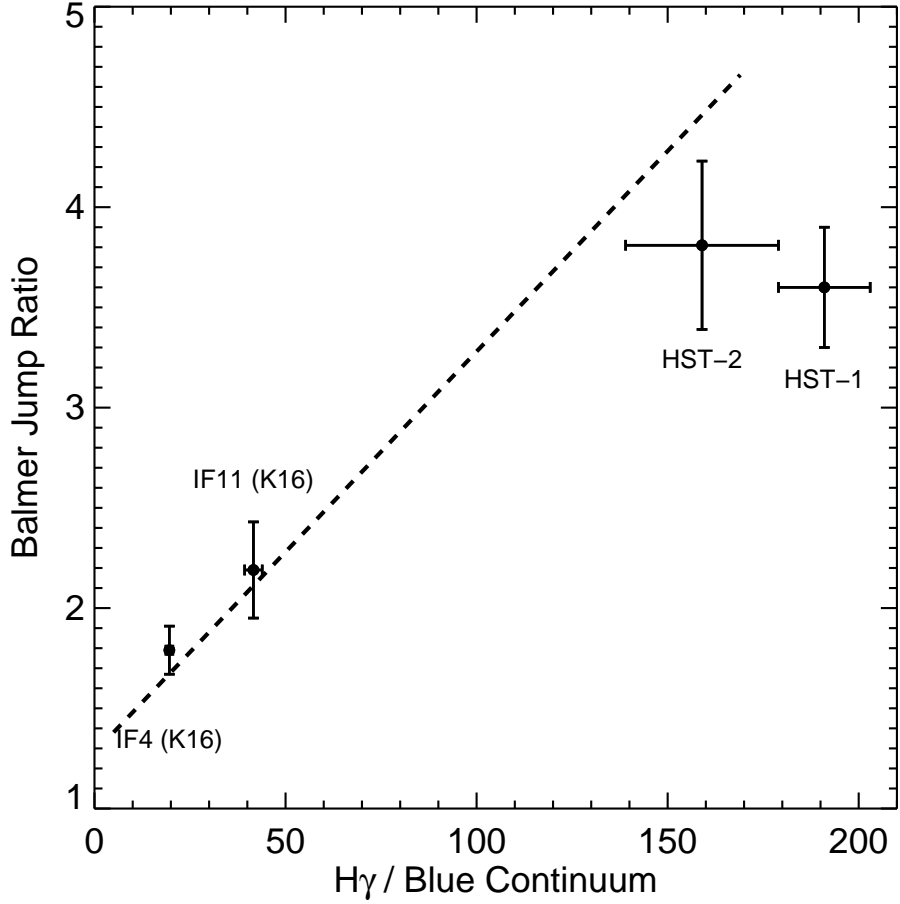}
\caption{ Large Balmer jumps ($\chi_{\rm{flare, peak}}$) occur in flares with more prominent Balmer line radiation (relative to the blue continuum flux, C4170\prim); the fit from K13 is shown as the \emph{dashed line}.  The values for HST-1, HST-2, IF4 (from K16), and IF11 (from K16) are consistent with this general trend, but HST-1 and HST-2 exhibit among the largest impulsive phase values of these quantities compared to the flares in K13. The error bars calculated as for the APO/DIS spectra of K13.}
\end{center}
\label{fig:hgamma}
\end{figure}

The spectral properties of the HST-1 and HST-2 events are similar to the characteristics of the GF-type events in K13, while the impulsiveness indices from the broadband photometry are similar to the HF-type events in K13.  Hereafter, we refer to HST-1 and HST-2 as HF/GF-type events.  
Over the flare peak times of HST-1 and HST-2, the moderate Balmer jump ratios ($\gtrsim 3.5-4$), large H$\gamma$/C4170\prim ratios ($140-200$), relatively large fraction of total flare flux due to Balmer radiation (0.3), large values of BaC3615/C3615\prim (0.7), and small impulsiveness indices (1.5) from the $U$-band light curves (Section \ref{sec:photanalysis}) are consistent
with properties of HF and GF-type events in K13.  These values are given in Table \ref{table:data} along with the Balmer line decrements. Notably, the values of H$\gamma$/C4170\prim\ ratios are among the largest observed in GF-type dMe flares in K13; as noted for another HF/GF type event on GJ 1243 in \citet{Silverberg2016}, this quantity is a useful proxy for the Balmer jump ratio.  We extend the linear relationship from K13 (cf. their Figure 11 and Equation 5) for the Balmer jump and line-to-continuum ratios in Figure \ref{fig:hgamma}; HST-1 and HST-2 flares establish a new regime among these quantities.

\begin{figure}
\begin{center}
\includegraphics[scale=0.70]{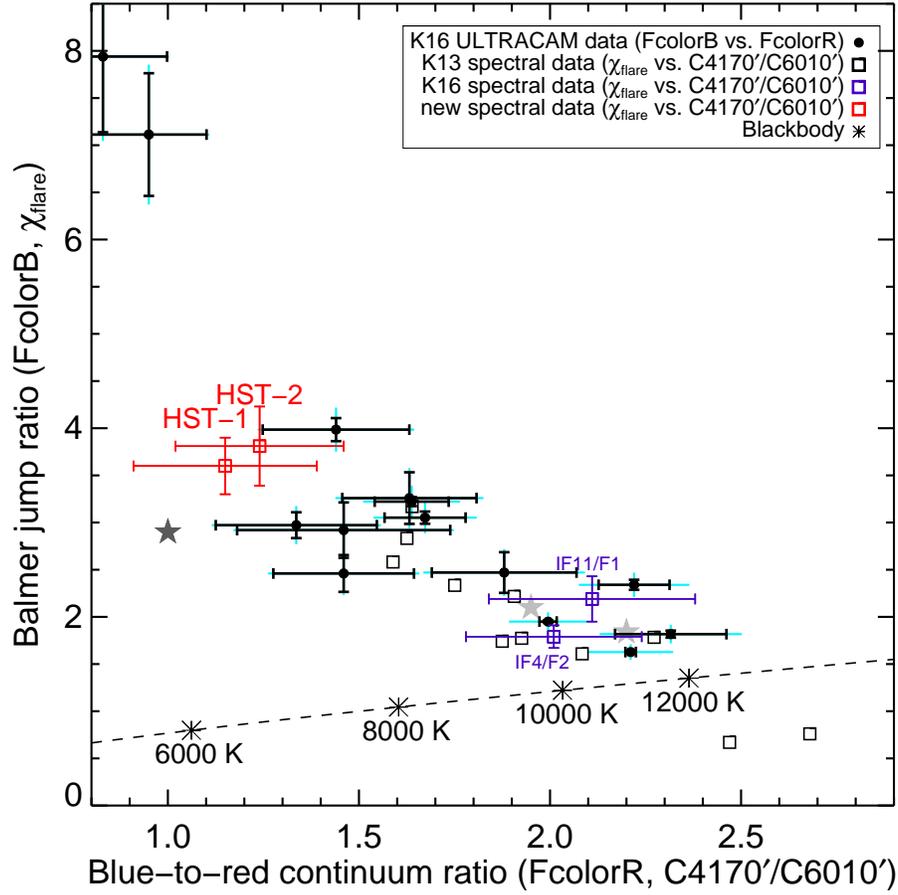}
\caption{ Color-color diagram of flares integrated over the peak times.  \emph{Squares} show values obtained from low spectral resolution spectra. HST-1 and HST-2 establish a new regime that are neither consistent with F11 RHD models (Figure \ref{fig:balmerjump}), hot blackbody curves, the F13 RHD models from K16 (\emph{light gray stars}), or other dMe flares that are more impulsive in their broadband time evolution.  Two similar amplitude but more impulsive flares (with ARC 3.5-m/DIS spectra) on YZ CMi from K16 are shown here as \emph{purple squares}:  IF4 at (2.01$\pm0.23$,1.79$\pm0.12$) and IF11 at (2.11$\pm0.27$, 2.19$\pm$0.24).   The values for HST-1 and HST-2 are given in Table \ref{table:data}.  The error bars on the flare colors from APO/DIS spectra are calculated as in K13.  For the flare-peak ULTRACAM colors,  the \emph{black} error bars show only the statistical uncertainties (for comparisons of flares on the same star and same observing night; see K16) while the \emph{cyan} error bars include a 5\% systematic uncertainty in the flare color calibration (see K13, K16).   The \emph{dark gray star} is the ``DG CVn superflare multithread (F13) model'' from \citet{Osten2016}; see text.  }
\end{center}
\label{fig:flarecolors}
\end{figure}

The HST-1 and HST-2 flare-only spectra exhibit moderate Balmer jumps similar to other HF and GF events in K13, placing them at one end of the empirical color-color sequence from K16.
The Balmer jump ratios of HST-1 and HST-2 indicate significant Balmer continuum radiation present, although there is suggestive evidence for $T\sim 8500-9500$ K blackbody radiation in the blue-optical (Table \ref{table:data}, Figure \ref{fig:balmerjump}) at a low flux level in HST-2.  However, the blue-to-red optical continuum distribution (C4170\prim/C6010\prim) is best fit by a lower temperature ($T_{\rm{FcolorR}}$, which is the blackbody color temperature that fits the ratio C4170\prim/C6010\prim; K16) by several thousand degrees.  The lower color temperature over wider spectral range may be due to a dominant, redder continuum component that has been referred to as the ``conundruum'' radiation in K13.  In Figure \ref{fig:flarecolors}, we show the Balmer jump ratio vs. blue-to-red optical flux ratio covering the peak times of HST-1 (S\#163) and HST-2 (S\#152-153).  The flare colors for HST-1 and HST-2 are statistically similar; averaging several spectra in the impulsive phase of HST-1 does not significantly change the location of this event in this color-color space.  These events fall to the upper, left region of the distribution for dMe flare peaks from K16, and their robust values (obtained from spectra) establish an unexpected new regime of continuum flux ratios that also includes two flares on YZ CMi (IF12 and GF1) observed with narrowband photometry from ULTRACAM (K16).   The colors of blackbodies are shown.   The divergence of the actual flare emission from blackbody colors reflects the increasing prominence of the Balmer jump and Balmer lines; a similar argument has recently been applied to coarse measurements of the Balmer jump in solar flare data in \citet{Hao2017}.

Many IF events and other HF events more clearly exhibit 10,000 K blackbody-like radiation at blue-optical wavelengths (cf. Figure 8 of K13).  The spectra of the HST-1 and HST-2 events are significantly different from the two moderate-amplitude events IF4 and IF11 on YZ CMi from K16.  The flare colors from 60~s integration time, ARC 3.5-m/DIS spectra of IF4 and IF11 from K16 are also indicated in Figure \ref{fig:flarecolors}.  These events lie much closer to the blackbody line at the other end of the color-color distribution with smaller Balmer jump ratios ($1.8-2.3$), hotter blue-optical continua
with optical color temperatures of $\sim9000-12,000$ K robustly constrained from spectra and from ULTRACAM photometry (see K16).  The more impulsive events have strikingly smaller line-to-continuum ratios as well (Figure \ref{fig:hgamma}).  Because IF4 and IF11 exhibit similar peak flux enhancements compared to HST-1 and HST-2, the amplitude of a flare in the $U$-band alone does not determine where a flare is located along the color-color sequence and thus if a flare produces energetically dominant $T\sim10,000$ K blackbody-like radiation.

\floattable
\begin{deluxetable}{lcccccccccccccc}
\rotate
\tabletypesize{\tiny}
\tablewidth{0pt}
\tablecaption{The $U$-band and Optical Peak Flare Spectral Properties}
\tablehead{
\colhead{Flare} &
\colhead{S\#} &
\colhead{UT} &
\colhead{HST orbit time [s]} &
\colhead{C3615\prim/C4170\prim\ (err)} &
\colhead{C4170\prim/C6010\prim\ (err)} &
\colhead{$T_{\rm{BB}}$ [K] (err) }&
\colhead{$T_{\rm{FcolorR}}$} &
\colhead{H$\gamma$/C4170\prim} &
\colhead{H$\gamma$/H$\delta$} &
\colhead{H$\gamma$/H$\beta$} &
\colhead{H$\gamma$/H$\alpha$} &
\colhead{H11/H$\gamma$} &
\colhead{BaC3615/C3615\prim} &
\colhead{HB frac} }
\startdata 
HST-1 & 163  & 00:46:16 - 00:47:16 & $2219 - 2279$  & 3.6 (0.3, 0.2)  & 1.15 (0.24,0.07) & 7300 (300) & 6500 (700)  & 191 (12, 5) & 1.1 & 0.8 & 1.0 & 0.15 & 0.72 & 0.48 \\ 
HST-2 & 152-153 & 07:06:38 - 07:08:50  & $2113 - 2246$   & 3.81 (0.42, 0.24)  & 1.24 (0.22, 0.10) & 9000 (500) & 5800-7200 & 159 (20, 5) &  1 & 0.9 &  1.23 & 0.1-0.14 & 0.71 & 0.46 \\
\enddata 
\tablecomments{The UT times occur on 2014-09-01.  See Figure \ref{fig:lcfigs}(a)-(b) for S\#\ numbering.  When two errors are given in parentheses, the first error corresponds to the result of error propagation when $\sigma$ is the standard deviation of flux values; the second error in parentheses corresponds to the result of error propagation when $\sigma$ is the error of the weighted mean.  The line-integrated flux of H11 was obtained by subtracting a constant determined as the flux between the H11 and H10 lines.  Prime symbols indicate flare-only specific flux values in continuum regions.  }
\end{deluxetable}\label{table:data}

\clearpage

\section{HST/COS Spectral Analysis} \label{sec:cosanalysis}
\subsection{Continuum and Line Identification}
\begin{figure}
\begin{center}
\includegraphics[scale=0.70]{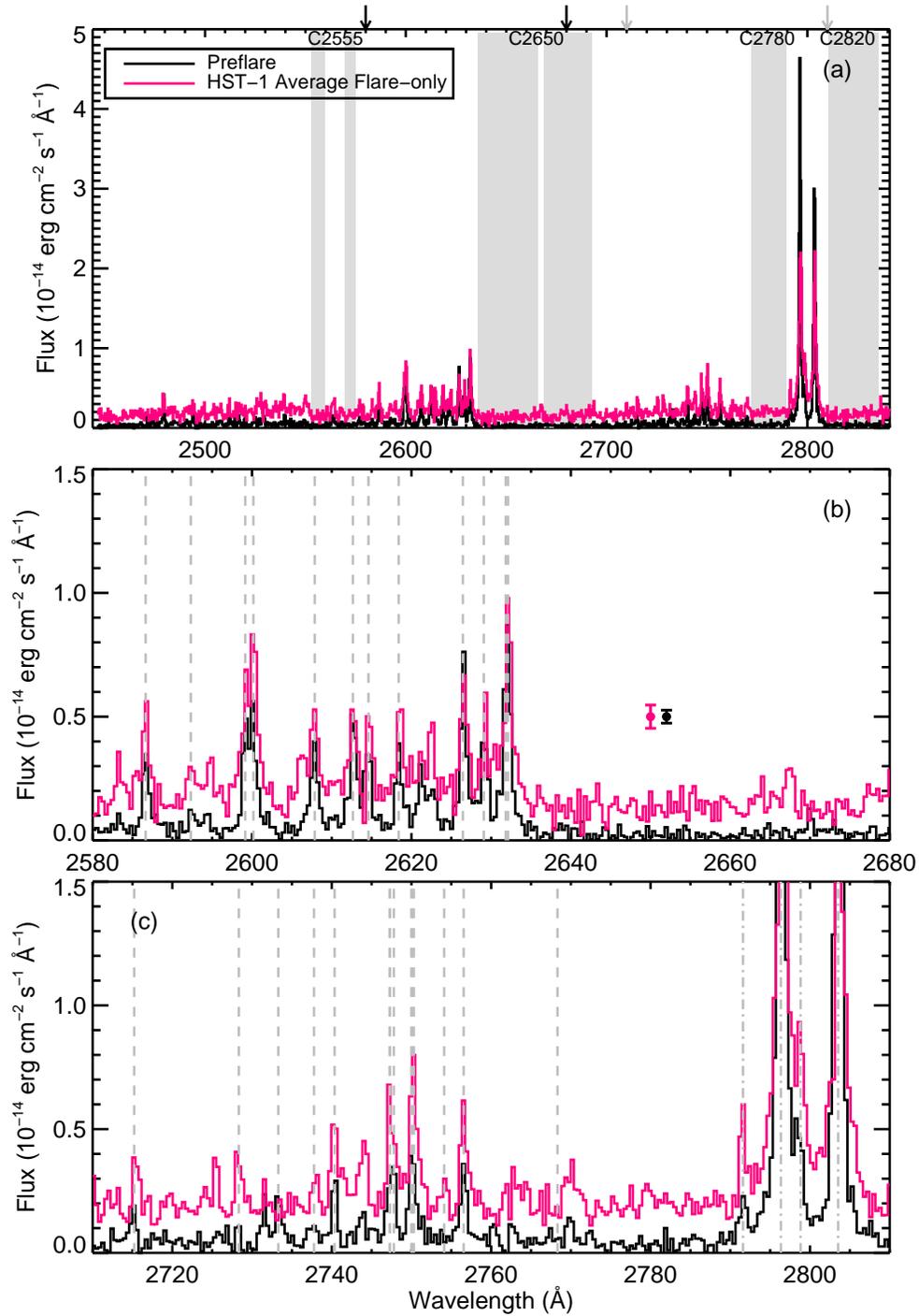}
\caption{ \textbf{(a)} HST-1 flare spectrum averaged over its duration, and a master pre-flare spectrum, are shown.   Filled grey shaded areas indicate continuum regions, black arrows indicate the 100 \AA\ expanded view in panel (b), and gray arrows indicate the 100 \AA\ expanded view in panel (c).  \textbf{(b)}  Representative error bars are shown, and a continuum enhancement is evident just redward of the Fe II of lines.  Rest wavelengths of Fe II lines from NIST are indicated.      \textbf{(c)} Dot-dashed lines indicate rest wavelengths of Mg II from NIST; the $k$ and $h$ lines are at 2796.35 \AA\ and 2803.53 \AA, respectively.  }
\end{center}
\label{fig:reference}
\end{figure}

We extract an HST/COS spectrum averaged over each flare's duration to identify continuum wavelength regions and emission lines.   Figure \ref{fig:reference}(a) shows master pre-flare and flare-only spectra for HST-1 over the NUV wavelength range $\lambda=2444-2841$ \AA\ (including the short-wavelength region affected by vignetting).
The pre-flare spectrum exhibits emission lines because the strong NUV lines form in the stellar chromosphere, where temperature generally increases outwards.
During the flare, the line-free wavelength windows from $\lambda=2635-2665$ \AA\ and $2668-2692$ \AA\ are indicated by grey rectangles; the average flare-only flux in these two windows is hereafter C2650\prim.  Three additional continuum windows that we use in the analysis are the following:  $2553.19 - 2559.84$ \& $2569.59 - 2574.7$ \AA\ (C2555\prim),
$2772.5 - 2788.5$ \AA\ (C2780\prim), and $2810.35 - 2834.92$ \AA\ (C2820\prim).  We calculate the average, standard deviation, weighted mean, and error in the weighted mean over each of these continuum windows.  
The flare produces bona-fide enhanced continuum 
flux in these windows that are line free above the noise (Figure \ref{fig:reference}(b)).  

Most of the wavelength range for the flare and quiescent spectra
consists of Fe II emission lines, where vacuum rest wavelengths from \citet{Nave} are retrieved from NIST and are indicated by vertical \emph{grey dashed lines}; many of these identifications have upper energy levels of $E_{\rm{upper}}/hc \sim40,000$ cm$^{-1}$, where $E_{\rm{upper}}$ is the upper level energy above the ground state of Fe II.  Models of Fe II will presented in a future work to constrain flaring temperatures and densities implied by this value of $E_{\rm{upper}}$.  We note that Fe II lines with $E_{\rm{upper}}/hc \sim 60,000$ cm$^{-1}$ are prominent in IRIS spectra  \citep{DePontieu2014} of solar flares in the NUV and these lines (in LTE) are sensitive to temperatures around $T\sim 8,000 - 18,000$ K for a range of densities \citep[][Kowalski et al. 2018, in prep]{Kowalski2017A}.  

The brightest flare Fe II lines in the HST spectra are the $\lambda2599.15+2600.17$ (hereafter, Fe II $\lambda2600$) and $\lambda2631.83+2632.11$ (hereafter, Fe II $\lambda2632$). 
We calculate the line-integrated, continuum-subtracted flux in these four Fe II lines and add them for a line-to-continuum ratio 
of Fe II / C2650\prim, similar to H$\gamma$/C4170\prim\ in the optical. 
The Mg II $h$ and $k$ lines are indicated by vertical \emph{dot-dashed lines} in panel (c).  
The Mg II triplet lines at 2791.6 \AA\ and 2798.8 \AA\
are also detected in and outside the flares and are indicated by \emph{dot-dashed lines} in (c).   We sum the Mg II line emission for the ratio of Mg II / C2650\prim.

\begin{figure}
\begin{center}
\includegraphics[scale=0.5]{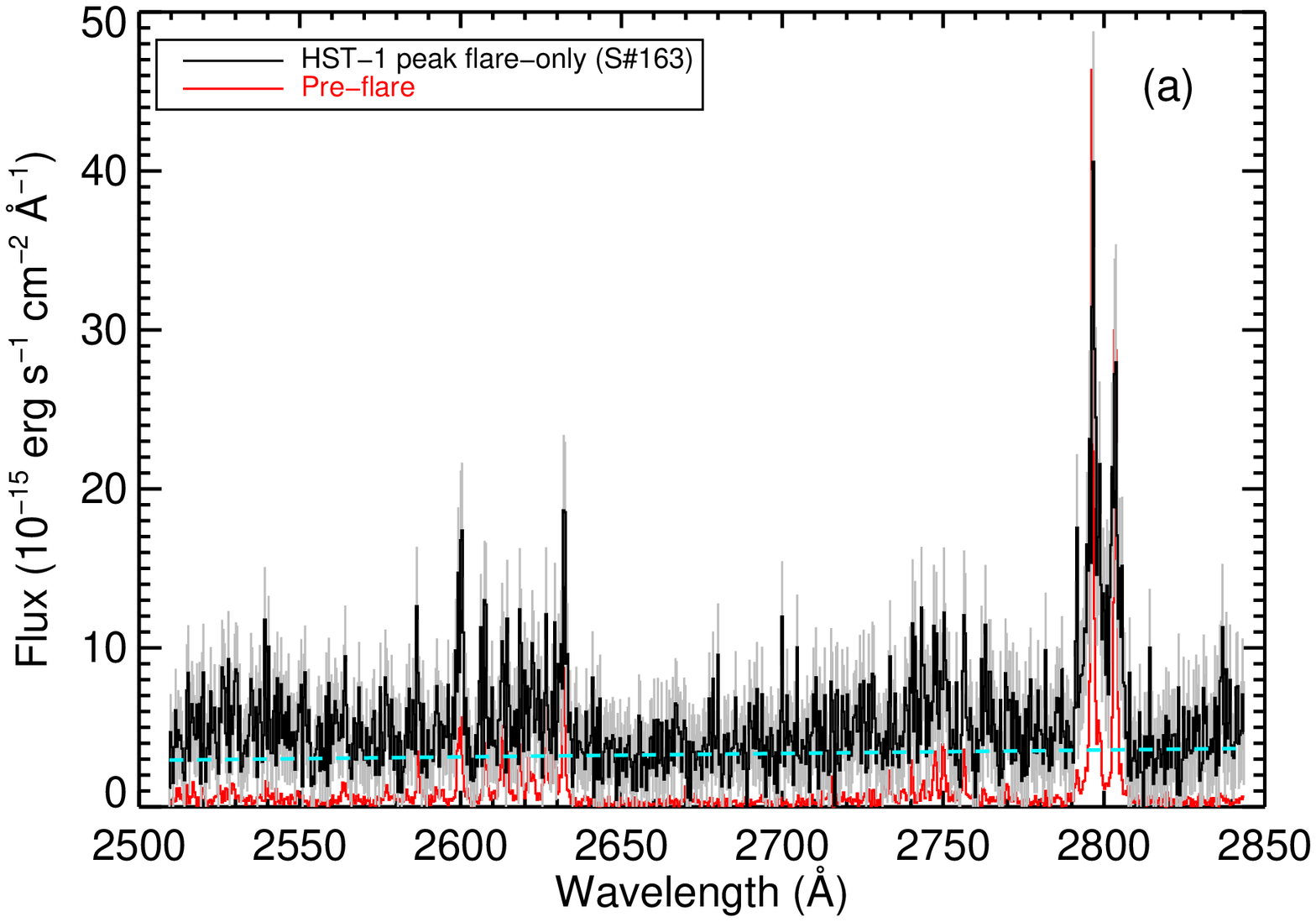}
\includegraphics[scale=0.5]{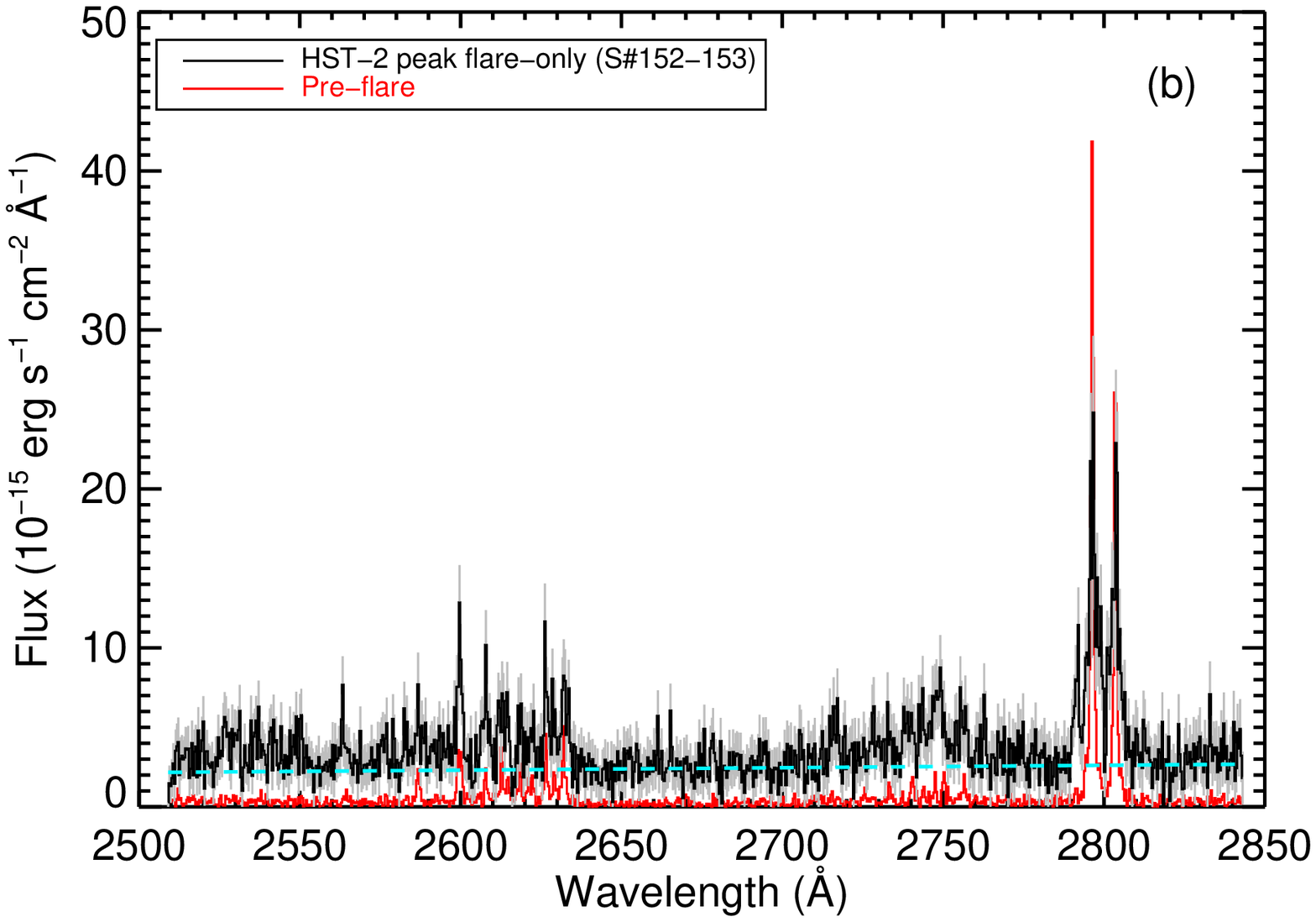}
\caption{ \textbf{(a)} HST/COS flare spectrum extracted over the time of S\#163 from the WHT ground-based spectrum.  \textbf{(b)} HST/COS flare spectrum extracted over the time of S\#152-153 from the APO ground-based spectra.   The best fit line (m$\lambda+$b) is shown in \emph{cyan} and is fit to the continuum windows indicated in Figure \ref{fig:reference}(a).  }
\end{center}
\label{fig:hst_peak_spec}
\end{figure}

\subsection{Peak COS Spectra Analysis} \label{sec:cos_peak_analy}
In Figure \ref{fig:hst_peak_spec}, we show the TIME-TAG spectra from HST/COS extracted over the times of S\#163 of HST-1 and S\#152-153 of HST-2, for which the optical and Balmer jump spectra were presented in Section \ref{sec:bj}.
The signal-to-noise is clearly low:   C2650\prim/$\sigma_{\rm{C}2650} = 2.2$ in panel (a) if we use the standard deviation of the flux over this wavelength interval as $\sigma$.  However,  the error in the weighted mean over the C2650\prim\ window gives a signal-to-noise of 16.  The signal-to-noise is 24 using the weighted mean of the four continuum windows, C2555\prim, C2650\prim, C2780\prim, and C2820\prim.

The slope of the continuum using a linear fit of the form $m\lambda+b$ to these four continuum windows helps constrain the value of $\lambda_{\rm{peak}}$ and the relative line and continuum energy in the flare.  These linear fits are nearly flat and are shown (\emph{cyan}) in Figure \ref{fig:hst_peak_spec} for HST-1 and HST-2.  For HST-1, we integrate under this best-fit line to find that $\sim60$\% of the flare energy from $2510-2841$ \AA\ at this time is due 
to the fitted continuum level.  The C2650\prim/C2820\prim\ NUV flare color is 0.94$\pm0.11$, indicating a rather flat continuum though with nearly 10\% uncertainty.  For HST-2, we integrate under the best-fit line to the continuum to find that 65\% of the flare energy from 2510-2841 \AA\ at this time is due 
to the continuum radiation.  The C2650\prim/C2820\prim\ flare color value is 0.88$\pm0.08$ for HST-2, which is also statistically consistent with an approximately flat continuum.  

These HST/COS spectra include the peak times with significant decay phase radiation (Figure \ref{fig:lcfigs}).  Therefore, we extract the TIME-TAG data of HST-1 over the rise phase, approximately corresponding to the times of S\#162 in Figure \ref{fig:lcfigs}.   We confirm that there is no significant difference in the linear fit to the continuum windows in this spectrum, and the flare color  C2650\prim/C2820\prim$=0.79\pm0.08$.

In Figures \ref{fig:hst_peak_spec} (a)-(b) the flare spectra qualitatively look as if they are scaled quiescent spectra.  However, scaling the quiescent spectrum to the value of C2650\prim\ reveals that the ratio of line-to-continuum flux in the quiescent is clearly much larger than in the flare.  This property is similar to optical flare spectra, in which many of the same very prominent lines are in emission compared to quiescence but the relative energy in the continuum greatly increases. 
Detailed modeling of HST flare continuum formation will be explored in a future work;  in these models, the explosive continuum enhancement is a result of electron beam heating that increases the optical depths, ionization fractions, and Balmer continuum emissivity at large chromospheric column mass.

 Two more differences are notable in the flare spectra compared to quiescence.  During the peak HST-1 and HST-2 spectra, the Fe II line complex at 2510-2565 \AA\ becomes stronger than in quiescence relative to the 2590-2635 \AA\ complex.  This may be expected since the 2590-2635 \AA\ complex is more optically thick, which may have a smaller relative increase in flux as the optical depths increase during the flare.   While this effect occurs in both HST-1 and HST-2, the relative flux in Fe II $\lambda$2632 Fe II $+\lambda2600$  to the Fe II $\lambda$2626.5 line (which shares the same upper state as the Fe II $\lambda$2600 line) differs: this ratio appears larger in HST-1 than in HST-2 and in quiescence.  These effects should be investigated with higher signal-to-noise data.

\clearpage
\subsection{Time Evolution}
In the optical, the emission lines decay slower than the continuum flux \citep{HP91}.  We speculate that this well-established phenomenon may be due to a decreasing energy deposition rate into the lower atmosphere throughout the flare, thereby causing the optically thinnest transitions (e.g., continuum, high-$n$ hydrogen lines) to fade by a larger percentage (relative to peak) as the heating at high column mass becomes weaker in the decay phase.  The HST-1 and HST-2 flare data allow us to characterize the Balmer continuum evolution at lower optical depth than at the Balmer series limit while also constraining the NUV Mg II and Fe II lines, which may evolve gradually like the Ca II K lines since calcium, iron, and magnesium have similarly low first ionization potentials of $6-8$ eV.  In the gradual phase of the Great Flare of AD Leo, the Mg II lines evolved similarly to the Ca II K line.    Differences in the time evolution would thus reveal properties of the optical depths in the same flaring plasma.  In Section \ref{sec:nuvdiff}, we compared the broadband NUV to the $U$-band photometry evolution at high cadence.  Here, we follow K13 and calculate the $t_{1/2}$ values and the peak-normalized light curves to compare the time-evolution of each emission line and continuum measure at lower time resolution, $t_{\rm{exp}} \sim60$~s. 

\begin{figure}
\begin{center}
\includegraphics[scale=0.35]{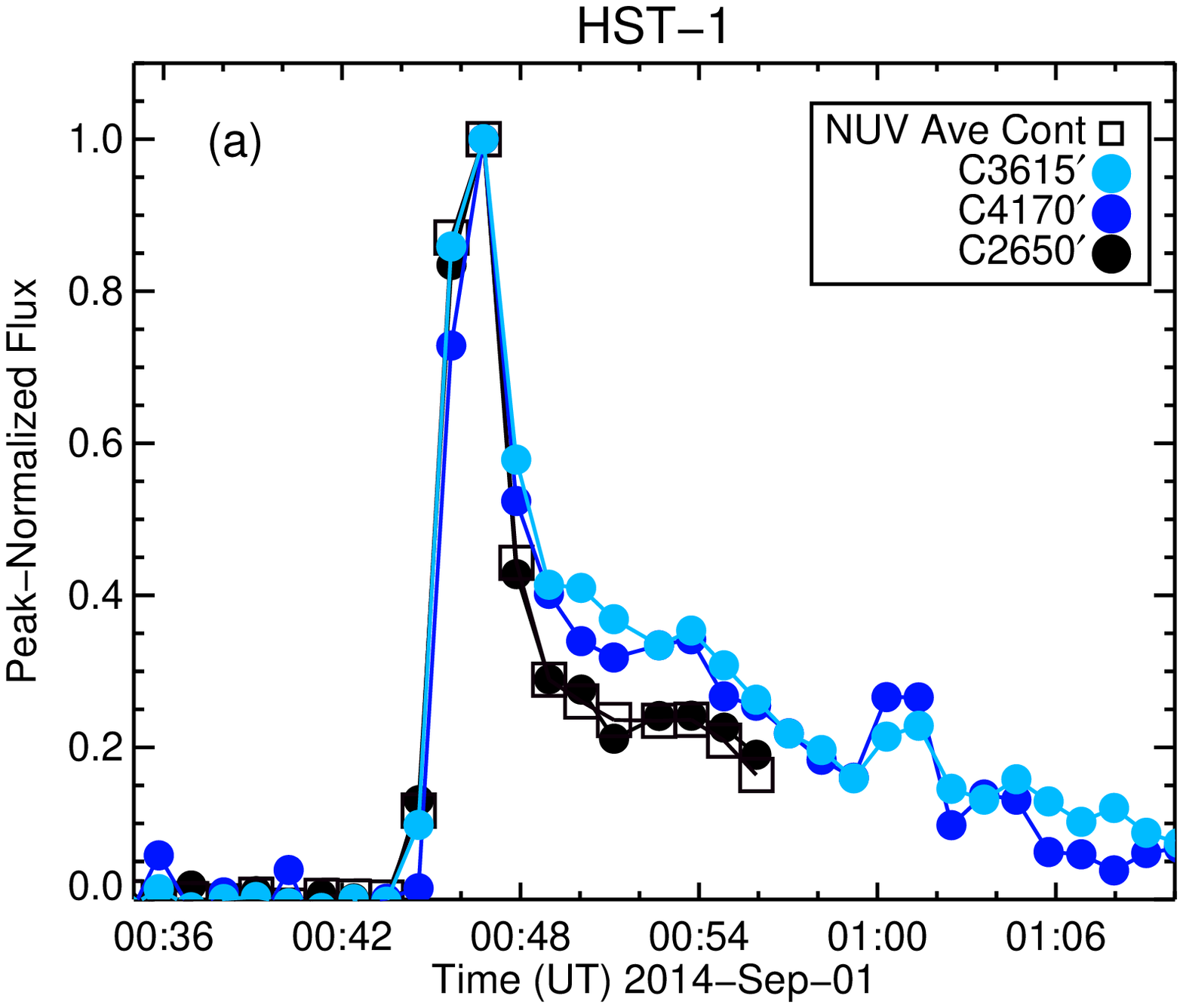}
\includegraphics[scale=0.35]{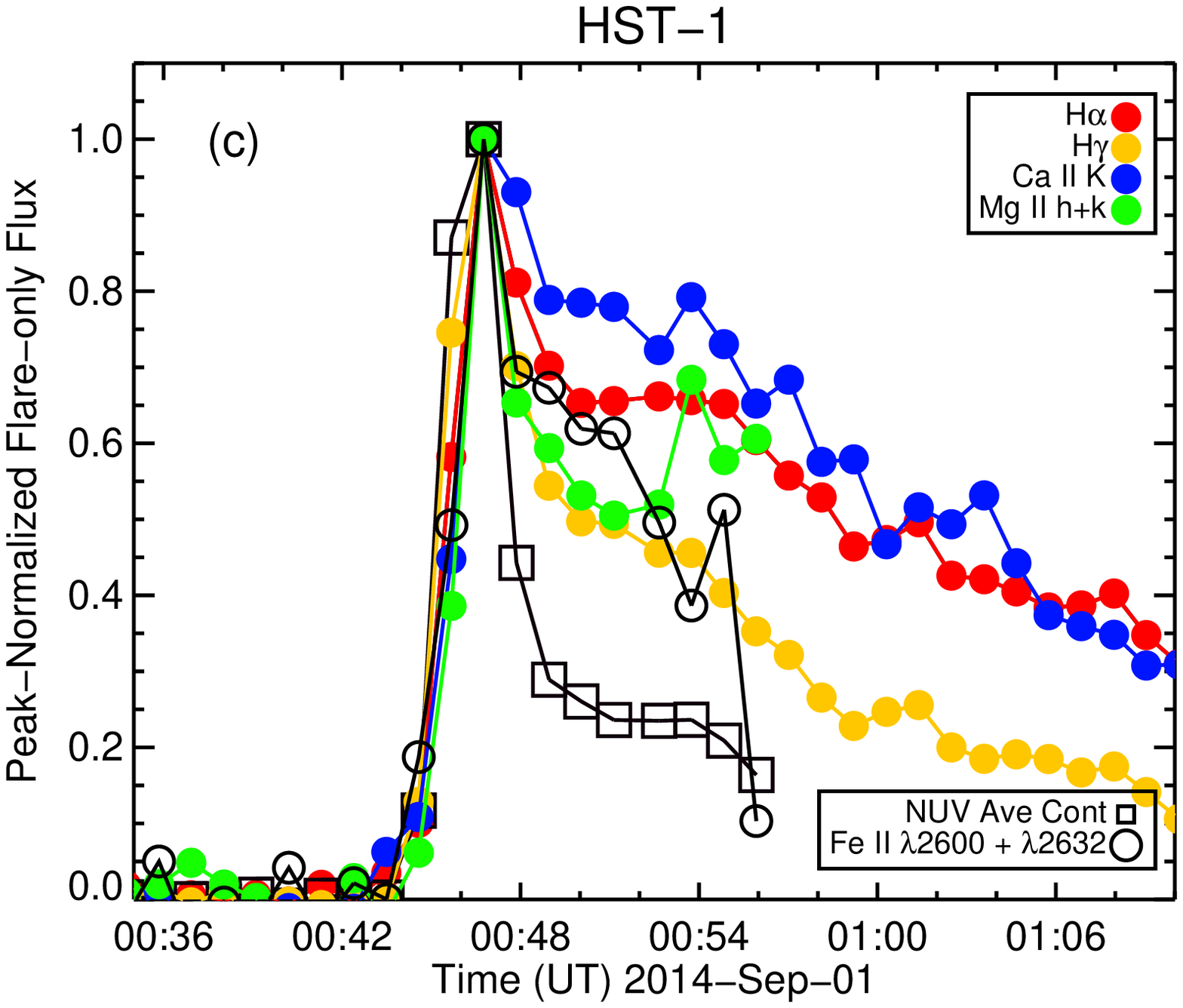}
\includegraphics[scale=0.35]{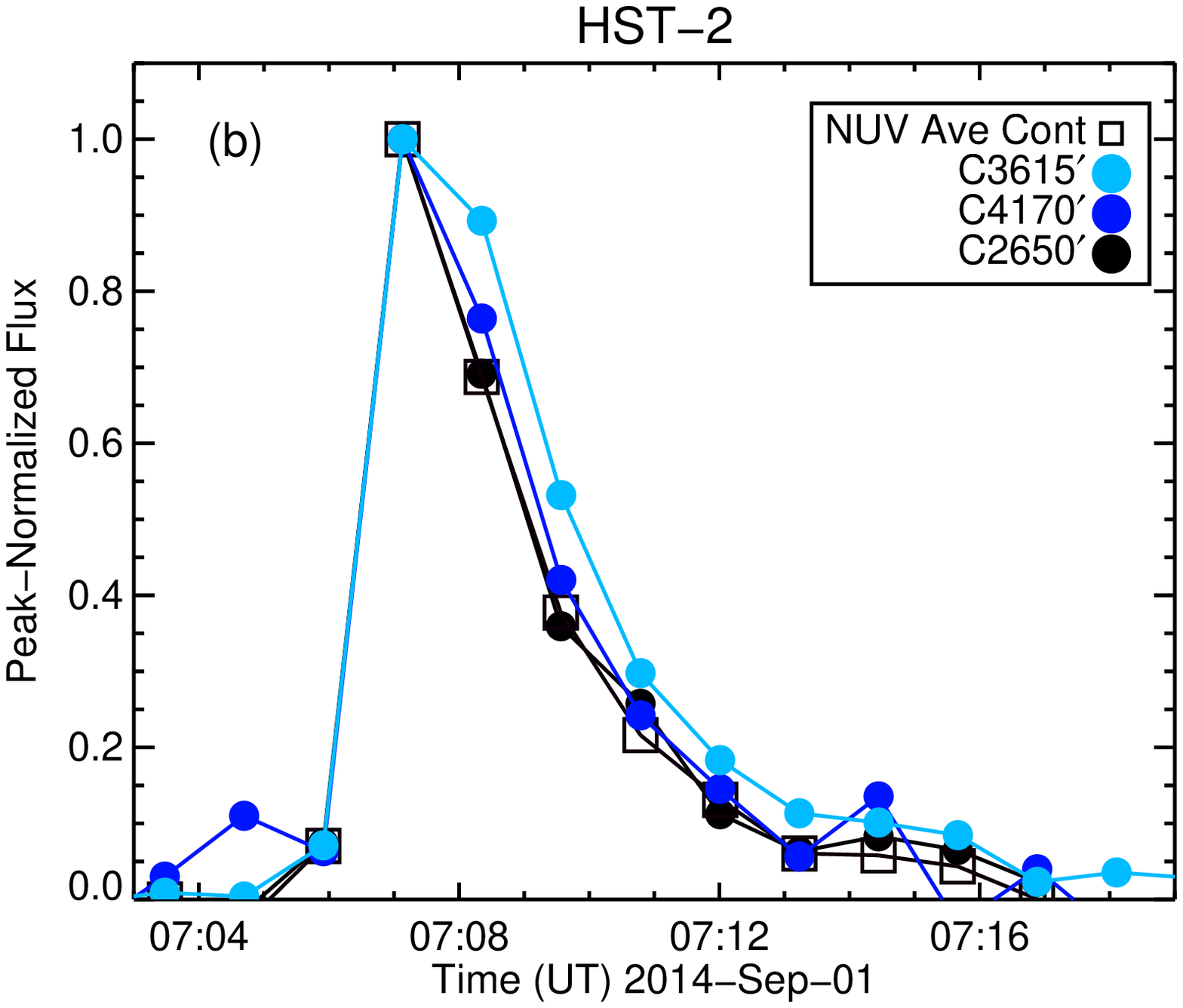}
\includegraphics[scale=0.35]{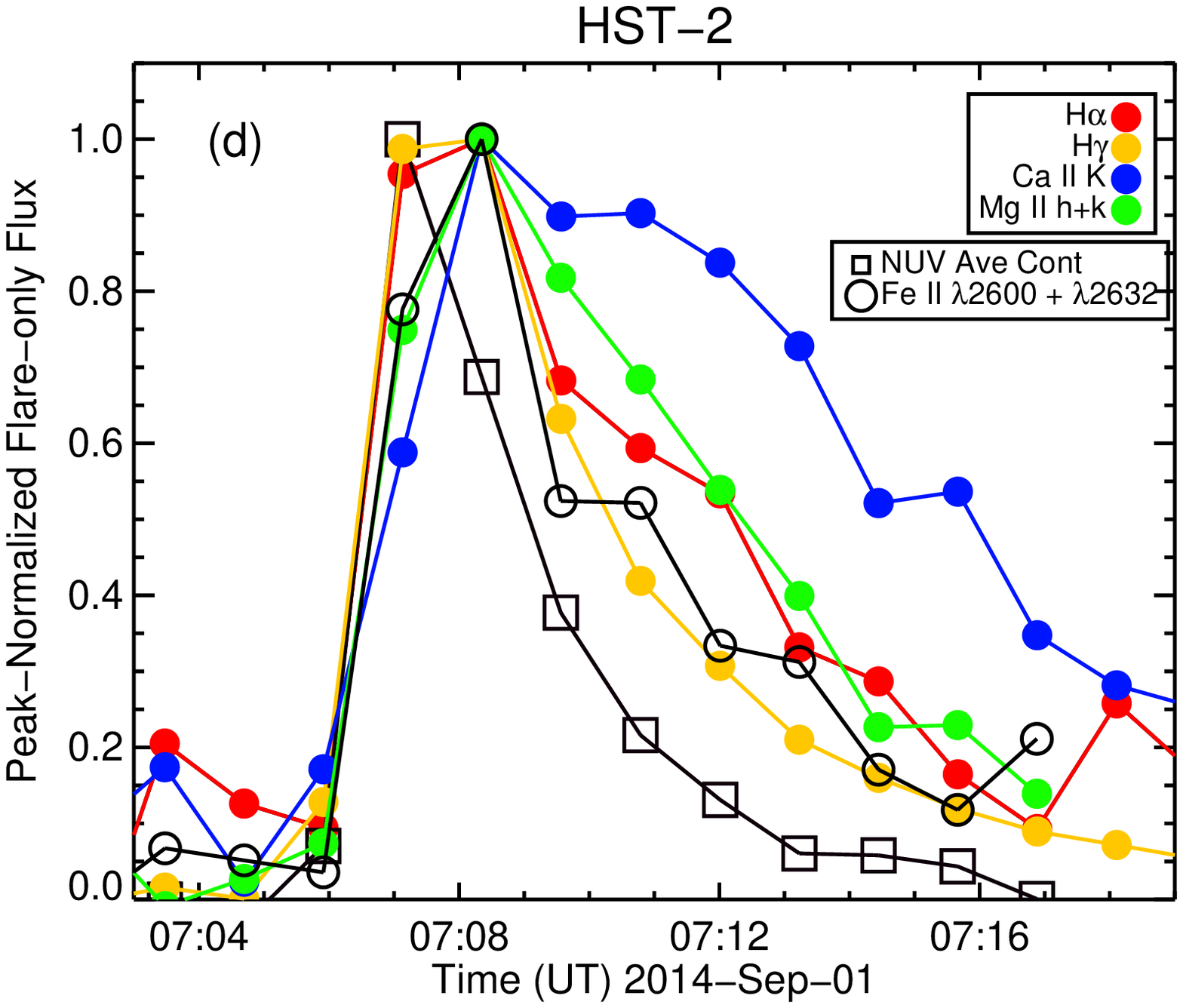}
\caption{ Peak-normalized time-evolution of continuum windows (a)-(b) and emission lines (c)-(d) for HST-1 shown over 35 minutes and HST-2 shown over 16 minutes.  For both HST-1 and HST-2, the NUV continuum evolution is faster than the emission lines, most notably the Fe II lines.  All continuum measures evolve similarly.  }
\end{center}
\label{fig:panels}
\end{figure}

We extract the COS/G230L TIME-TAG data over the duration of each ground-based spectral exposure time for HST-1 and HST-2 in Figure \ref{fig:lcfigs} and subtract the master pre-flare spectrum for each flare.
In Figure \ref{fig:panels}, we show the time evolution of C2650\prim\ compared to the ground based-spectral continuum evolution of 
C3615\prim\ and C4170\prim\ for HST-1 (panel a) and HST-2 (panel b).  The weighted-mean of the four HST continuum windows (``NUV ave cont'') is also shown to closely follow C2650\prim, as expected.  The evolution of the continuum windows over the NUV is similar\footnote{The continuum light curves in the NUV in Figure \ref{fig:lcfigs} seem to show a faster decay than the C3615\prim\ and C4170\prim, but this difference is not significant when taking the un-weighted mean of the NUV continuum windows.} to the evolution of C3615\prim\ and C4170\prim\ at low time resolution. 
 Compared to the emission lines (panels (c)-(d)), it is clear that the NUV continuum windows exhibit a closer time-evolution
to the Balmer continuum (C3615\prim) and the blue-optical continuum (C4170\prim).

Figure \ref{fig:panels}(c) shows the peak-normalized, line-integrated flux evolution for HST-1 for several emission lines.  The Mg II flux is similar to H$\gamma$ flux in the decay, but it weakly responds to the first impulsive heating event like Ca II K.
HST-1 exhibits the typical pattern in the decay among spectral quantities: from most elevated relative to peak to least elevated, the ordering is the following:  Ca II K, H$\alpha$, and higher order Balmer lines (e.g., H$\gamma$) which are all slow compared to the relative decline of the continuum measures, similar to other flares in K13.
The Fe II lines in the HST spectra are slower to decline than the HST NUV continuum in HST-1.  
The time-evolution differences between the NUV continuum measures and Fe II in the HST spectra, and the similarity among the panchromatic continuum measures, suggest that the NUV continuum measures are not misidentified from a blend of faint, Fe II lines that are unresolved at the relatively low-spectral resolution of these observations.   In HST-2 (Figure \ref{fig:panels}(d)), Mg II has a slower rise and decay than all lines except for Ca II K.  Here, the peak-normalized time-evolution of each Balmer line is similar, which has been noted for other HF events in K13;  in IF events, the higher order hydrogen lines are faster to fade than the lower order hydrogen lines, such as H$\alpha$.

\begin{figure}
\begin{center}
\includegraphics[scale=0.5]{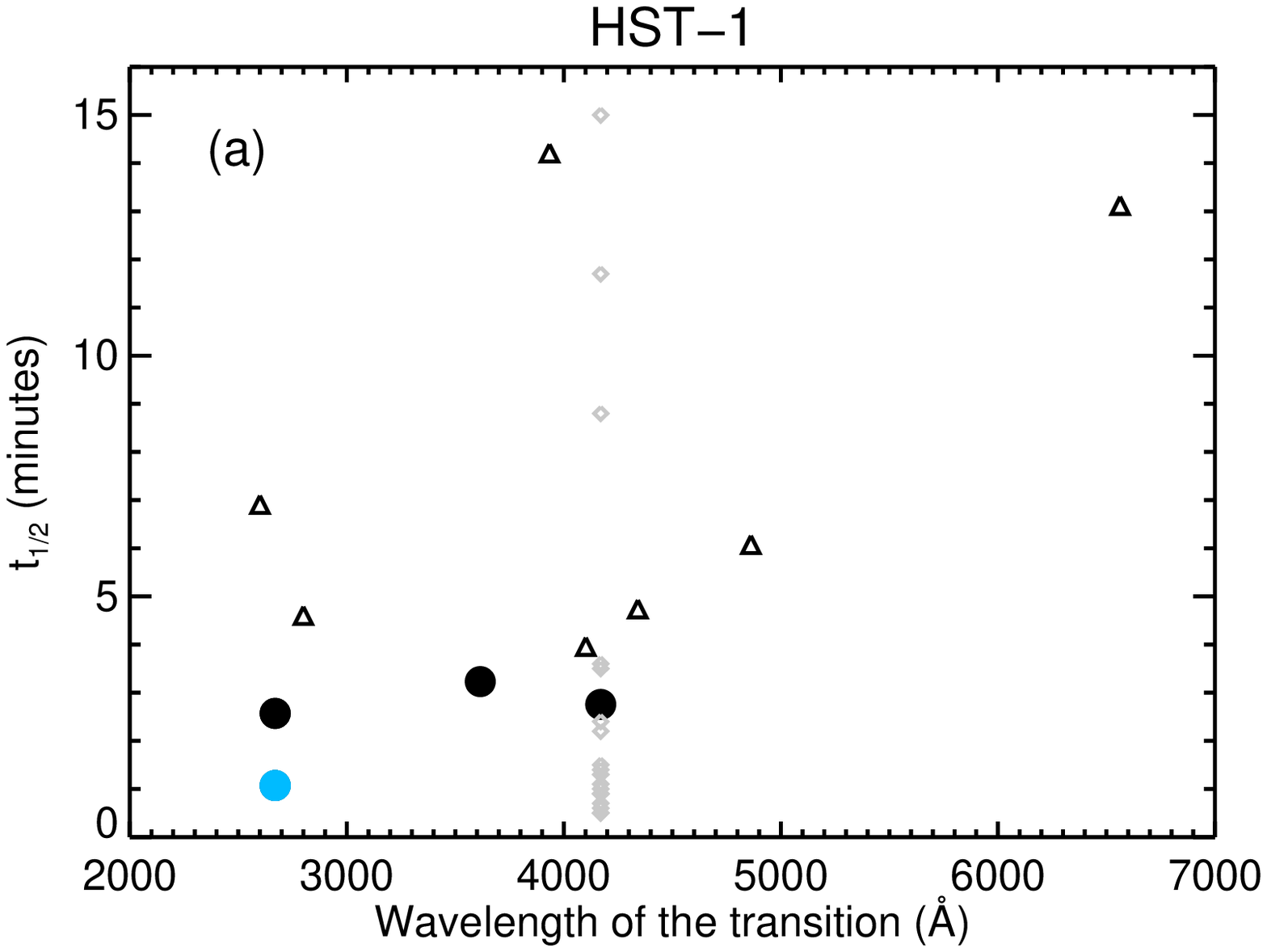}
\includegraphics[scale=0.5]{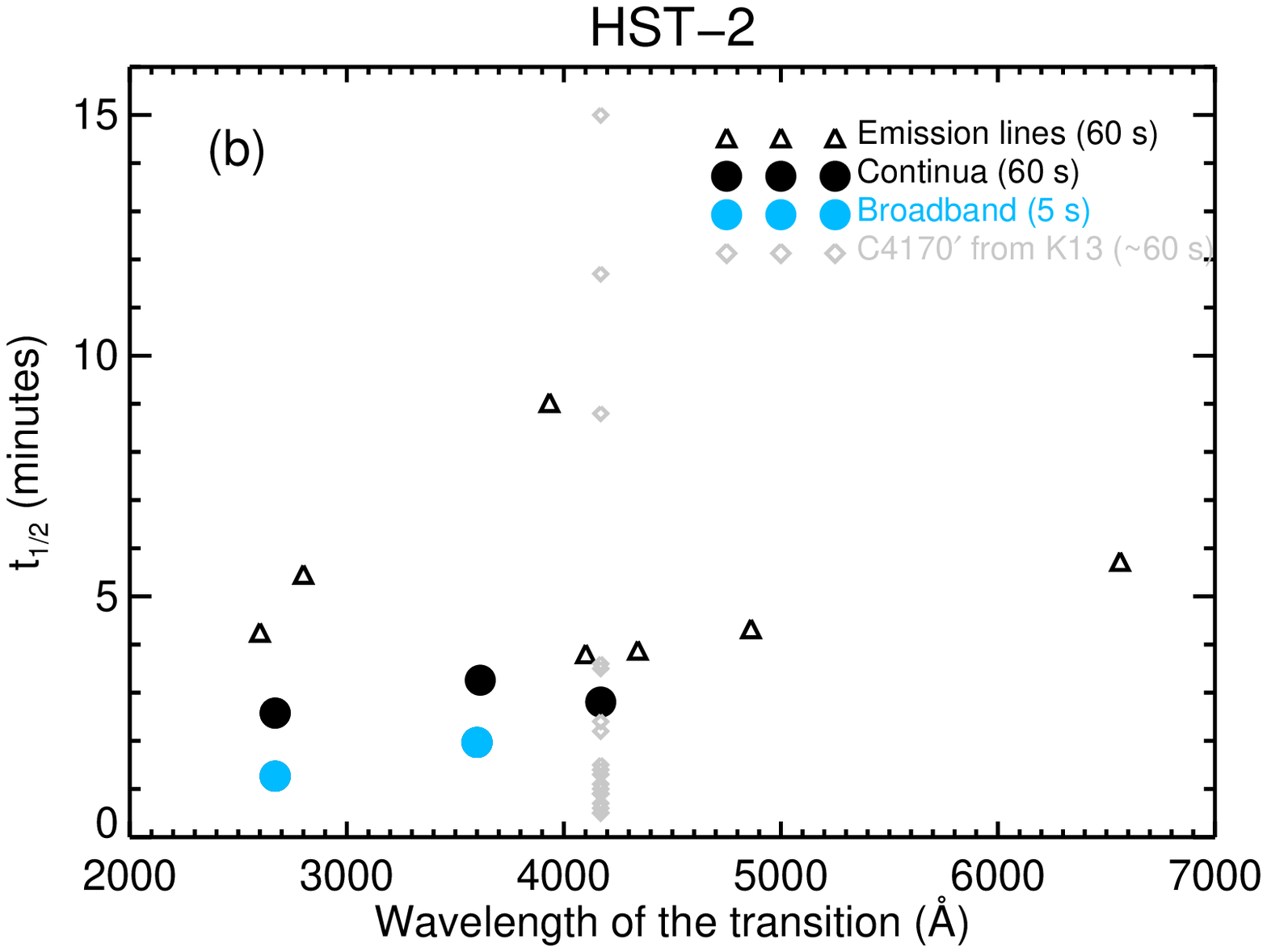}
\caption{  The time-decrements for HST-1 and HST-2 calculated from the peak-normalized light curves in Figure \ref{fig:lcfigs}; the FWHM of the light curve is the $t_{1/2}$ value.  Emission lines are indicated by \emph{triangles}, continuum measures C2650\prim, C3615\prim, and C4170\prim\ as \emph{filled black circles}, and the broadband (NUV and $U$-band) measures at high-time ($t_{\rm{exp}}=5$~s) resolution by \emph{blue filled circles}.  The continuum and broadband measures are the fastest quantities.  The \emph{grey diamonds} are values of C4170\prim\ from other flares  (at similar temporal $\sim30-60$~s resolution) from K13.  The apparent impulsive phases (Figure \ref{fig:lcfigs}, \emph{teal asterisks}) of HST-1 and HST-2 are relatively long-duration at this time-resolution due to several shorter periods of fast and gradually rising emission that are evident 
in the high-time ($t_{\rm{exp}}=5$~s) resolution light curves (Figures \ref{fig:lcfigs} - \ref{fig:uband}).   }
\end{center}
\label{fig:thalf}
\end{figure}

We also plot the $t_{1/2}$ value of each spectral component against the wavelength of the transition (following K13), which is known as a time-decrement, for HST-1 and HST-2 in Figure \ref{fig:thalf}.
HST-1 exhibits a time-decrement like some other IF-type events from K13, but the H$\alpha$ value is larger than a linear extrapolation from the higher order lines, as occurs in the GF-type time-decrement presented in K13 (the classification as ``hybrid'' is thus appropriate).    HST-2 has a flat time-decrement like the HF-type event discussed in K13.  The $t_{1/2}$ values of the C4170\prim\ quantity are larger than many values calculated from similar temporal resolution data in K13 (\emph{grey diamonds}).

From these data of HST-1 and HST-2, we conclude that Mg II does not rise as fast as the Balmer lines, and is more like Ca II K, while it decays like one of the lower-order Balmer lines, such as H$\alpha$ or H$\gamma$.   Among all the emission lines in these events, the Ca II K line has the most gradual time-evolution. The Fe II lines have a rise like Mg II but a decay like that of H$\gamma$.  The Fe II lines are not good proxies of the NUV continuum measures in these two flares; perhaps these relatively strong Fe II lines  are too optically thick to originate from the same deep layers where continuum radiation can escape.  The Fe II / C2650\prim\ values covering the peak times of HST-1 and HST-2 are given in Table \ref{table:hstdata}.  The time-evolution of the emission lines will be compared in detail to RHD models to test the hypothesis that the different decay rates are due to variation in the optical depths of the lines.   

Emission line profile changes and line shifts would also provide strong constraints on the radiative-hydrodynamic modeling \citep{Kuridze2015, Brown2018}.  The HST spectra exhibit anomalous mid-orbit dispersion shifts by one pixel (Section \ref{sec:cosdata}), which prevents us from readily constraining the time-evolution of the line shifts using the techniques for EUV and NUV solar flare spectra \citep{Graham2015, Brown2016}.  Recent observations of a long-duration flare on the dM3.5e star EV Lac showed striking blue wing enhancements in H$\alpha$ over several hours \citep{Honda2018}. The spectral resolution in the optical data of GJ 1243 is too low to robustly characterize line shifts and asymmetries, but we compare the H$\alpha$ equivalent width evolution of HST-1 (from the WHT/ISIS red spectra with 10~s exposure times) to this EV Lac flare.  The HST-1 event achieves a peak equivalent width of 14 \AA\ in H$\alpha$ with a rise time of only two minutes, while the EV Lac flare has a much longer rise time ($\sim10-30$ minutes) to a peak equivalent width of $\le$11 \AA.  We speculate that these two flares may correspond to different regimes of explosive and gradual chromospheric heating rates \citep[e.g.,][]{Fisher1989}.

\subsection{Flare Data at the $U$-band Atmospheric Cutoff (3120 - 3500 \AA)} \label{sec:keck_lris}
\begin{figure}
\begin{center}
\includegraphics[scale=0.5]{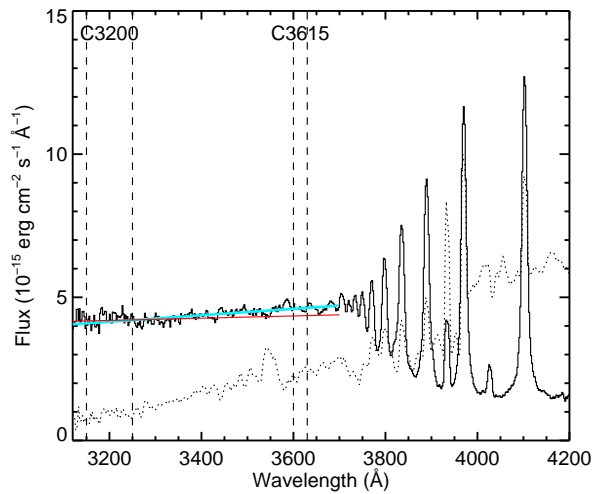}
\caption{ Keck/LRIS spectra from 07:06:48 UT to 07:08:55 UT, covering the HST-2 peak and corresponding to similar times of the spectrum in Figure \ref{fig:balmerjump}.  A linear fit to 3120 - 3700 \AA\ is shown as the cyan line, and the pre-flare from Keck is the dotted line.   The red line is a fit to 3430 - 3700 \AA\ in Figure \ref{fig:balmerjump} and extrapolated to shorter wavelengths.
The slopes and flux agree remarkably well.  Arrows in Figure \ref{fig:lcfigs}(b) indicate the times over which two Keck/LRIS spectra are averaged.    }
\end{center}
\label{fig:lris}
\end{figure}

The C3200\prim/C3615\prim\ is a new flare color spectral index that is obtained from the Keck/LRIS spectra, which have wavelength coverage at $\lambda>3120$ \AA.  We average over S\#116 (07:06:48 - 07:07:33) and S\#117 (07:08:10 - 07:08:55) from the Keck/LRIS flare-only spectra of HST-2 and show this spectrum\footnote{For HST-1, we do not calculate this quantity because it is not possible to continuously rotate the slit at the parallactic angle at the WHT.} in Figure \ref{fig:lris}.   We fit a line to the wavelengths from 3120 \AA\ to 3700 \AA.  The value of C3200\prim/C3615\prim\ is 0.93$\pm0.05$, which is consistent with the ratio obtained from the linear fit. We also fit a line to $\lambda=3430-3700$ \AA\ in the APO spectrum of HST-2 from Figure \ref{fig:balmerjump} and extrapolate to compare to the slope obtained from Keck.   The linear fit is slightly flatter in the APO flare spectrum, which could be due to flux calibration uncertainties in the Keck/LRIS (see Section \ref{sec:keckdata}).  Generally, the absolute flux values and slopes are in good agreement (Figure \ref{fig:lris}).

\section{New Constraints for RHD Models} \label{sec:combined}
How do the short wavelength HST/COS and Keck/LRIS spectra supplement the longer-wavelength NUV and optical spectral constraints on RHD models?
 Since we do not have spectra of GJ 1243 with overlapping wavelength coverage, we evaluate the fidelity of the absolute flux calibration of the HST/COS data in order to compare to the optical flux.  The spectra from the ground-based telescopes are calibrated to the $V-$ and/or $B-$band magnitudes of GJ 1243 (Section \ref{sec:apodata}). The
spectrophotometric evolution is consistent with the $U$-band photometry (Figure \ref{fig:uband}) for HST-2, while the broadband enhancement in HST-1 in the red is consistent with the Aristarchos $V+R$ photometry (Figure \ref{fig:aristarchos}).  We follow \citet{Sirianni2005} and calculate the filter-weighted specific flux density of the pre-flare NUV spectrum of HST-2 using the effective area response matrix of Swift/UVOT/UVW1, which has a central wavelength at $\sim2600$ \AA.  We use a serendipitous Swift/UVW1 observation of GJ 1243 retrieved from MAST (Obs ID 00040116002; $t_{\rm{exp}}=232.6$~s) and an aperture radius of 1.2x the PSF FWHM to obtain a count rate.  This count rate is converted to specific flux density using the conversion provided by the Swift team.   Compared to the Swift/UVOT photometry, the filter-weighted specific flux density of the HST-2 pre-flare spectrum is 7\% lower.  We confirmed that the pre-flare continuum level for HST-1 is consistent with the pre-flare continuum of HST-2.  By inspection of the spectra, the emission lines in HST-1 are elevated by nearly a factor of two over the pre-flare fluxes of HST-2; therefore the synthetic UVW1 flux is sensibly 30\%\ higher since there is a larger fraction of energy in the emission lines in quiescence in the NUV \citep[e.g.,][]{Hawley2007}.  At the peaks of HST-1 (S\#163) and HST-2 (S\#152-153), we extract HST/COS spectra using an aperture of $\pm57$ pixels and divide by the spectra extracted with the smaller, higher signal-to-noise aperture of $\pm15$ pixels (see Section \ref{sec:cosdata}), thus obtaining an aperture correction of 14$\pm8$\%, which is consistent in both flares.  In this section, we apply this aperture correction as a scale factor (1.14) to the HST/COS spectra to compare to the flux-calibrated spectra from the ground-based telescopes.  The quoted absolute calibration accuracy of G230L is somewhat larger than 2\% (COS ISR 2010-01), and thus we adopt a conservative uncertainty of 10\%\ for the absolute flux calibration of the HST/COS spectra.  

\begin{figure}
\begin{center}
\includegraphics[scale=0.70]{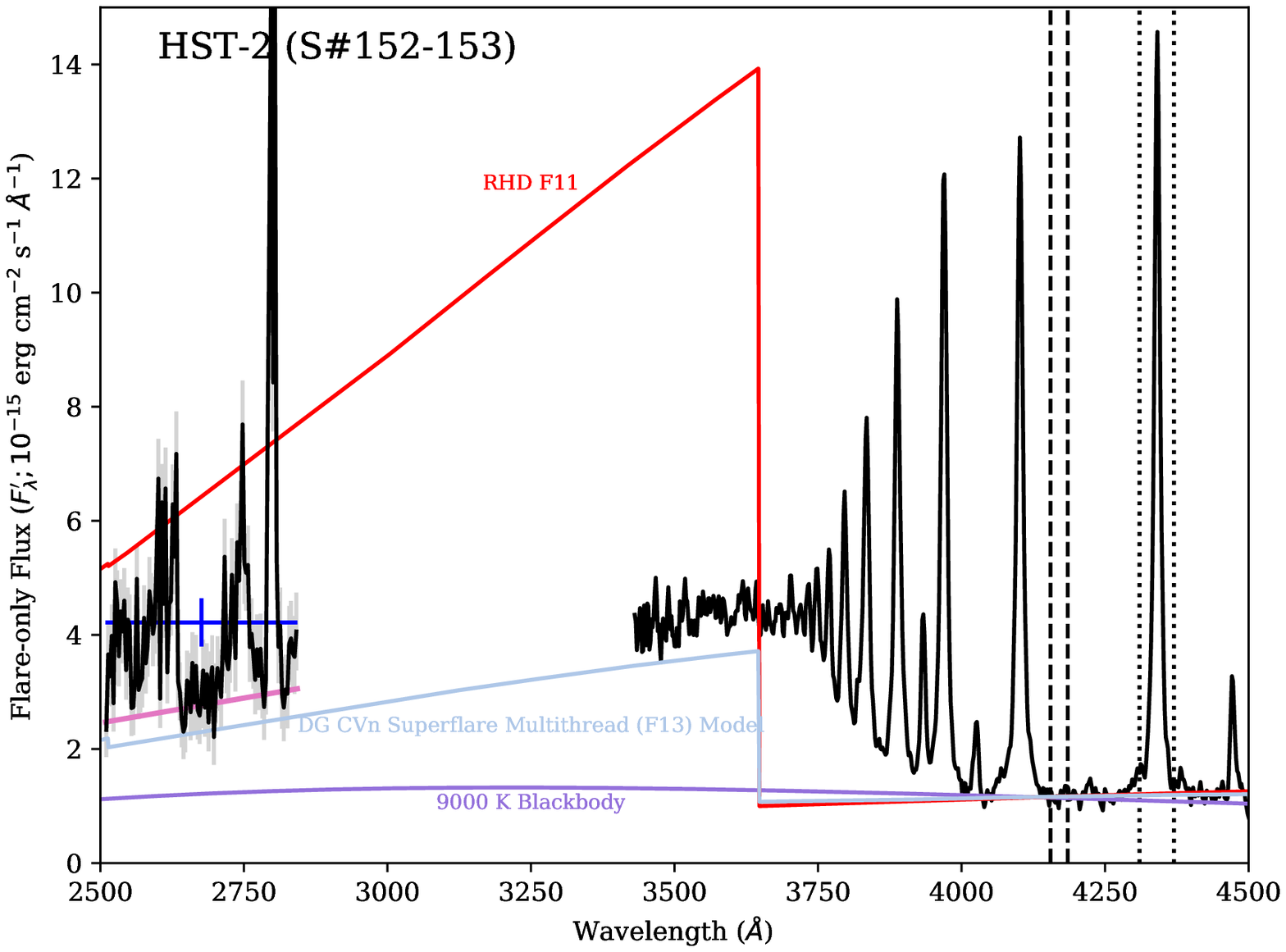}
\includegraphics[scale=0.70]{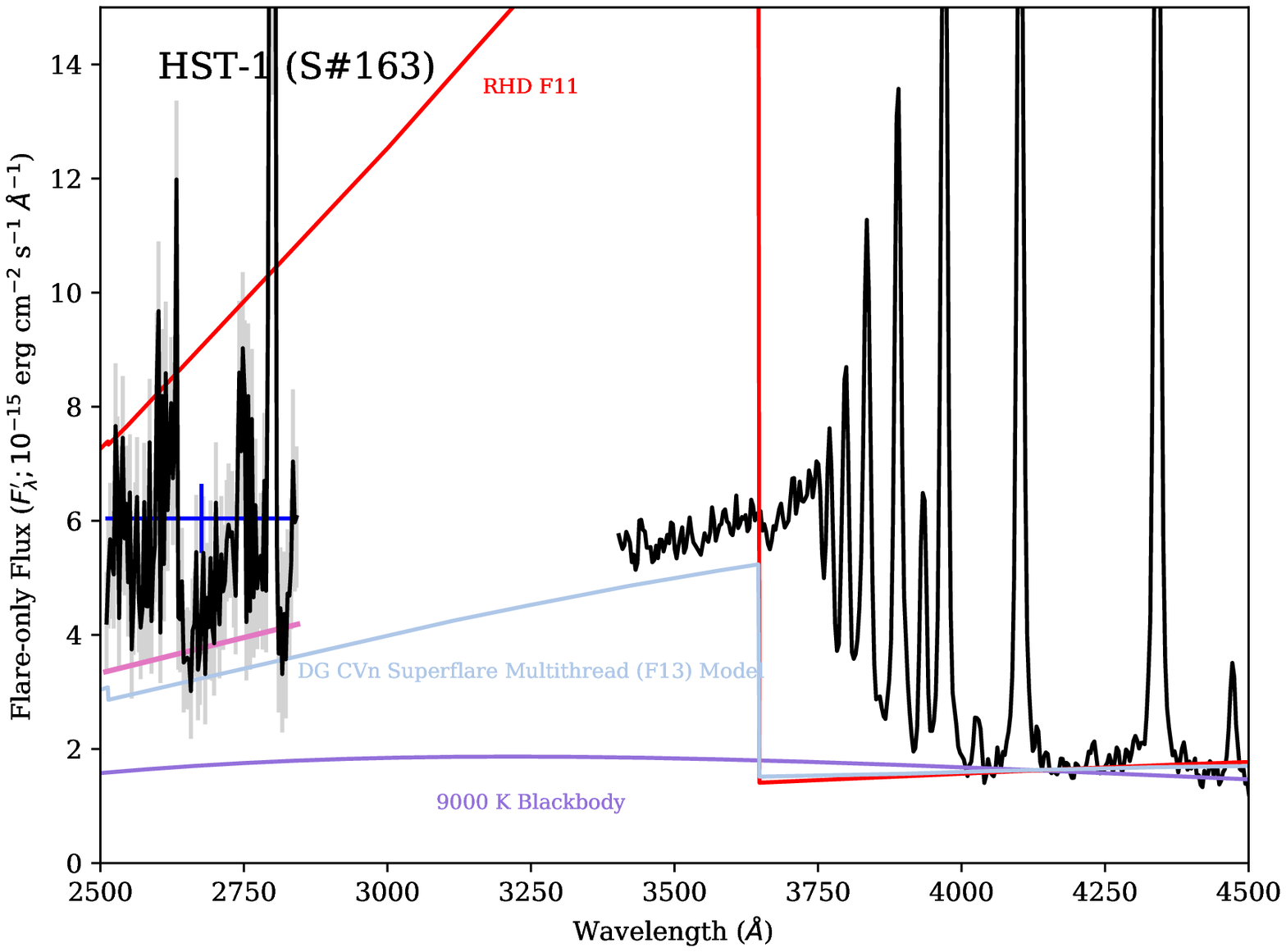}
\caption{ \textbf{(a)} Data of HST-2 for S\#152-153 in Figure \ref{fig:balmerjump} and HST/COS spectra binned to the same time interval and similar spectral resolution.  The wavelength ranges for C4170\prim\ and for the H$\gamma$ line-integrated flux are indicated by vertical \emph{dashed lines} and \emph{dotted lines}, respectively.  Three continuum models with a range of Balmer jump ratios are shown, normalized to the specific flux density of the APO spectra at $\lambda=4170$ \AA\ (the blackbody is taken from Figure \ref{fig:balmerjump} and is fit to the blue-optical continuum windows).  The \emph{pink line} is the continuum fit from Figure \ref{fig:hst_peak_spec}(b), and the \emph{dark blue point} shows the flare excess over the NUV wavelength range with an error bar indicating a (conservative)10\% systematic uncertainty in the absolute flux calibration of HST; see text.  \textbf{(b)} Same as in panel (a) but showing HST-1 corresponding to the times of S\#163 from the WHT.  Note that the optical continuum in this flare is rather flat (Figure \ref{fig:fullsed}); here we scale a $T=9000$ K blackbody to the C4170\prim\ of this spectrum, which satisfactorily matches the excess continuum shape between $4000-4400$ \AA. \label{fig:ultimate} }
\end{center}
\end{figure}

Figure \ref{fig:ultimate} shows the composite NUV spectra for S\#152-153 during HST-2 (panel a) and S\#163 during HST-1 (panel b).   The HST/COS data are shown at approximately the same spectral resolution.  These pan-chromatic flare spectra demonstrate that a $T=9000$ K blackbody that is extrapolated from the optical does not satisfactorily 
explain the NUV flare continuum over the impulsive phases of these HF/GF-type events.   In Figure \ref{fig:ultimate}, we show the average  $\lambda=2510-2841$ \AA\ specific flux density (\emph{blue points}), including line and continuum radiation.  This broadband flare-only specific flux value is approximately equal to the Balmer continuum flare-only specific flux in the middle of the $U$-band (e.g., C3615\prim).  The 9000 K blackbody underestimates the continuum flare flux by a factor of two at the NUV wavelengths; the broadband flux including emission lines (which result in pseudo-continua in this range) is under-estimated by a factor of three compared to the blackbody extrapolation.

Figure \ref{fig:ultimate} also shows how two RHD models from the literature compare to the NUV data.  The F11 RHD model \citep[from][]{Kowalski2015} clearly exhibits too large of a Balmer jump.  If scaled to the C3615\prim\ value, the flux decrease towards short wavelengths in the Balmer continuum is consistent with the lower error bars of the underlying continuum in the NUV range.  However, the C4170\prim\ is vastly under-estimated as noted in Section \ref{sec:bj}.  Another problem with the F11 RHD model is the very high H$\gamma$/C4170\prim\ value.  Although  H$\gamma$/C4170\prim\  was reported to be 150 in \citet{Kowalski2015} (and thus consistent with these observations), the line calculations in that work employed a Voigt profile with the damping parameters from \citet{Sutton1978}  for the electric pressure broadening\footnote{Collisional broadening from ambient protons and electrons, which is also sometimes referred to as the linear Stark effect.}.  However, \citet{Kowalski2017B} implemented an accurate hydrogen broadening prescription \citep[as first pointed out for solar flare spectra in][]{JohnsKrull1997} using the RH code \citep{Uitenbroek2001} to model flare atmospheres.  We use the RH code to recalculate the F11 RHD flare spectrum \citep[at $t=2.2$~s; see][for details]{Kowalski2015}  using the new hydrogen line profiles (``TB09HM88'') from \citet{Tremblay2009}.  The new H$\gamma$/C4170\prim\ value is much larger near $\sim330$, which is far from the observed range of $150-200$ in HST-1 and HST-2.  As expected, the new electric pressure broadening profiles make the hydrogen Balmer lines broader and brighter \citep[see also][for details]{Kowalski2017B}.  These profiles are much more accurate in comparison to the absorption lines of Vega's spectrum or in white dwarf spectra and are thus preferred for modeling the flare broadening, which is due to large electron densities and optical depths.  

The F13 multithread continuum model (``DG CVn superflare multithread (F13) model'') from \citet{Osten2016} is shown in Figure \ref{fig:ultimate}.  The multithread model is an average burst spectrum of impulsive heating ($\Delta t=2.3$~s) and impulsive cooling ($\Delta t = 2.7$~s) from the F13 
model of \citet{Kowalski2015}, where impulsive cooling refers to the phase of relaxation after the electron beam is turned off at 2.3~s.   The multithread model also includes an additional snapshot at $t=4$~s to represent the gradual cooling of previously heated loops over a 25x larger area.  The multithread approach originates from modeling spatially unresolved soft X-ray light curves of solar flares \citep{Warren2006}.
 The DG CVn superflare multithread (F13) model  is most consistent with the continuum flux ratios, the line-to-continuum ratios \citep[line radiation not shown here; see][]{Kowalski2017B}, and the NUV, $U$-band, and blue-optical continuum flux distribution.  This model was recalculated with the improved hydrogen broadening prescription in \citet{Kowalski2017B};  
the value of H$\gamma$/C4170\prim\ is 120 \citep[Table 2 of][]{Kowalski2017B}, making it deficient in line radiation compared to the observations.

Each continuum (or line-to-continuum) flux ratio is compared to the models and given a score as $\frac{F_{\lambda,\rm{obs}}-F_{\lambda,\rm{model}}}{\sigma}$.  
In the NUV range, we use C2650\prim/C2820\prim, where the values of C2650\prim\ and C2820\prim\ are calculated using the weighted means and the errors of these weighted means.  
 We use consistently defined values of $\sigma$ for the colors from the ground-based data to compare to the ratios from HST.  Thus, we use the weighted mean and the error in the weighted mean in each continuum window for the score calculation, whereas the error bars on the flare colors of HST-1, HST-2, IF4, and IF11 in Figure \ref{fig:flarecolors} use the standard deviation of the flux in each continuum window (see K13 for justification) in the propagation of the uncertainty of the flare color.   Here, we also add in quadrature a systematic uncertainty in the color calibration for the ground-based spectra ($\sim5$\% for each flare color index; see Appendices of K13 and also K16).  This error propagation\footnote{Calculated in this manner, the flare color uncertainties for IF4/F2 and IF11/F1 in Figure \ref{fig:flarecolors} are $\sim0.1$.  Compared to K13, the error bars from APO/DIS spectra in Figure \ref{fig:flarecolors} are conservative estimates for these two events as well.}  gives 
$\chi_{\rm{flare}} = 3.81 \pm0.24$ and C4170\prim\ /C6010\prim$=1.24\pm0.1$ for HST-2 (S\#152-153).   These uncertainties are the second values in the parentheses in Table \ref{table:data} and are more consistent with the flare color error propagation for the peak-flare ULTRACAM photometry (\emph{cyan} error bars) in Figure \ref{fig:flarecolors} here.

Table \ref{table:modchk} shows the scores for the following flare colors of HST-2 (S\#152-153):  C2650\prim/C2820\prim$=0.88 \pm0.08$, C3200\prim/C3615\prim$=0.93\pm0.05$, and H$\gamma$/C4170\prim$=159 \pm 5$.  The ratios of C2650\prim/C3615\prim\ and C2650\prim/C4170\prim\ include a  10\% systematic uncertainty in the flux calibration (from the aperture correction above) of the HST/COS  spectra:  C2650\prim/C3615\prim$=0.59\pm0.07$ and C2650\prim/C4170\prim$=2.24\pm0.26$.   For HST-1,  these two ratios are similarly 0.6 and 2.2, respectively.  Thus, the observed Balmer continuum decreases by $\sim60$\%\ from the Balmer series limit to $\lambda \sim 2650$ \AA, constraining $\lambda_{\rm{peak}}$ to the $U$-band, likely near the Balmer series limit.  Coincidentally, some RHD model snapshots that reproduce a dominant hot blackbody-like spectrum (e.g., the F13 model at $t=2.2$~s) produce a value of $\lambda_{\rm{peak}}$ that is near $2650$ \AA.   

We quantitatively evaluate the models in Figure \ref{fig:ultimate} using a new RHD grading scheme, which is the sum
of the absolute values of the scores for each spectral flux ratio, divided by the number of flux ratios.  
The final score represents the average $\sigma$-difference per model comparison measure and is given for each model in the last column of Table \ref{table:modchk}.  The F11 RHD model and the 9000 K blackbody are $\gtrsim 10 \sigma$ different per measure, making them poor representations to the data of HST-2. Excluding C3200\prim/C3615\prim, the final scores are similar for HST-1 (S\#163):  3.5, 11.3, and 14.6 for the three respective models in Table \ref{table:modchk}.  The DG CVn superflare multithread (F13) model is a significantly better representation of the observed flare spectra in these HF/GF events.  

\clearpage
\begin{deluxetable}{lccccccccc}
\rotate
\tabletypesize{\tiny}
\tablecaption{Models vs. HST-2 (S\#152-153 times)}
\tablewidth{0pt} 
\tablehead{
\colhead{Model} &
\colhead{Source} &
\colhead{$\frac{\rm{C}2650\rm{'}}{\rm{C}2820\rm{'}}$} &
\colhead{$\frac{\rm{C}3200\rm{'}}{\rm{C}3615\rm{'}}$} &
\colhead{$\frac{\rm{C}3615\rm{'}}{\rm{C}4170\rm{'}}$} &
\colhead{$\frac{\rm{C}4170\rm{'}}{\rm{C}6010\rm{'}}$} &
\colhead{$\frac{H\gamma}{\rm{C}4170\rm{'}}$} &
\colhead{$\frac{\rm{C}2650\rm{'}}{\rm{C}3615\rm{'}}$} &
\colhead{$\frac{\rm{C}2650\rm{'}}{\rm{C}4170\rm{'}}$} &
\colhead{Final score}}
\startdata 
DG CVn Superflare Multithread (F13) Model & O16, K17, K15  & $-0.1$ & $+1.0$ & $+3.9$ & $+2.4$ & $+7.8$ & $-0.8$ & $+1.5$ & 2.5  \\
RHD F11 $t=2.2$~s                                                & K15         &  $+0.3$ & $+1.8$ & $-23.2$ & $+3.7$ & $-34.1$ & $-0.6$ & $-11.1$ & 10.7  \\
9000 K blackbody                                                   &  Planck     &    $-0.8$ & $-1.7$ & $+11.1$ & $-3.7$ & $-65.9$ & $-3.2$ & $+4.1$ & 12.9  \\ 
\enddata 
\tablecomments{The values in this table give the quantity $\frac{F_{\lambda,\rm{obs}}-F_{\lambda,\rm{model}}}{\sigma}$ for each model comparison. Positive values indicate that the slopes of the models are not blue enough or that there is not enough line radiation compared to continuum radiation for the H$\gamma$/C4170\prim.   All models were calculated with the RH code from snapshots from radiative-hydrodynamic simulations using the RADYN code \citep{Carlsson1997, Allred2015, Kowalski2015}, except the blackbody.  The error bars for the 9000 K blackbody include an uncertainty of 500 K on the blackbody temperature.  Note, the DG CVn Superflare Multithread (F13) Model has a Balmer jump of 2.9 with RH and 3.2 with RADYN.   The 9000 K blackbody score would be much worse than the RHD F11 model if the constraints from other Balmer emission lines are included.  O16 refers to \citet{Osten2016}, K17 refers to \citet{Kowalski2017B}, and K15 refers to \citet{Kowalski2015}.}
\end{deluxetable}\label{table:modchk}

\section{Discussion II:  Implications for Flare Heating in HF/GF-type events} \label{sec:discussion2}

The DG CVn superflare multithread model has the lowest (best) grade among the three models tested here.  
Future verification of this model for other HF/GF-type events has several interesting implications.  For example, bursts of F13 electron beam heating adequately reproduce the line and continuum radiation in HST-1 and HST-2, as in the decay phase of superflares and megaflares \citep{Osten2016, Kowalski2017B}.  The high-time cadence light curves of HST-1 and HST-2 show that the impulsive phase spectra include peak times as well as fast and gradual decay radiation when the blackbody temperature often decreases rapidly (K13).  Thus, the average of many flare bursts, or many ``threads'' (individual flare loops), at various times in their evolution 
is a sensible picture based on solar flare observations and modeling \citep{Warren2006}.   The hot blackbody-like radiation in the F13 only appears for a small fraction of time that the electron beam impulsively heats the atmosphere.  In the multithread (burst average) model, the hot blackbody-like radiation gets diluted, while the rising and decaying radiation in each flare thread dominates the spatially-integrated spectra.  This multithread model includes a rather ad-hoc continuum and emission line component (the $t=4$~s snapshot in the F13 model) that represents the gradual decay of previously heated loops long before the burst\footnote{Excluding this ad-hoc spectrum from the multithread model causes the line-to-continuum ratio to be 74 and the Balmer jump ratio to be smaller as well \citep{Kowalski2017B}.}.  Better gradual cooling phase RHD models are needed.  

The interpretation of HST-1 and HST-2 as spatially and temporally averaged F13 beam-heated loops suggests that hot blackbody radiation may exist in HF/GF-type events but is produced for only a very short time and over a very small area in the flare.  A hot blackbody function was fitted to the blue-optical in the YZ CMi Megaflare decay phase spectrum \citep{Kowalski2010}, but the fit to the data was recently improved with the DG CVn superflare multithread (F13) model in \citet{Kowalski2017B}.  However, the YZ CMi Megaflare decay phase spectrum exhibits\footnote{These spectral quantities become smaller in the Megaflare's secondary events, which are explained as Vega-like-emitting sources \citep[K13 and][]{Kowalski2017B}.} a smaller Balmer jump ratio (2.7) and line-to-continuum ratio (90) compared to HST-1 and HST-2.  Although the main times of the flares studied here (S\#163 times for HST-1 and S\#152-153 times for HST-2) include significant amounts of time when the broadband NUV radiation is decaying (Figure \ref{fig:lcfigs}), the Balmer jumps are moderately high ($3-4$) in any other ground-based spectrum over the impulsive phases of these flares.  Also, the HST spectrum extracted over the rise phase times of HST-1 (Section \ref{sec:cos_peak_analy}) does not show anything significantly different to suggest that hot blackbody radiation is produced.  

 Compared to HST-1 and HST-2, the HF events in K13 have notably smaller values of H$\gamma$/C4170\prim\ $<75$ while they also exhibit stronger evidence of hot blackbody, $T=10,000-12,000$ K continuum radiation from $\lambda=4000-4800$ \AA. 
An observed (qualitative) relationship between the Balmer line radiation and the red continuum radiation has been established in K13; it appears that HST-1 and HST-2 are consistent with this relationship by having a prominent, redder/flatter continuum which may provide insight into the so-called ``conundruum'' radiation when modeled in detail with RHD simulations.  We note that flat red-optical continua and large line-to-continuum ratios are also observed in the decay phase of giant flares (Figure 30 of K13) and megaflares (Figure 31 of K13).  We speculate that the hot blackbody-like radiation may be present in the HST-1 and HST-2 events, but a flatter and redder continuum and the Balmer continuum clearly dominate the spectra.  

\subsection{Future Modeling Directions \& Connection to dG Flares}
In Paper II, we will combine constraints from the continuum ratios, line-to-continuum ratios, and the blending of the high-$n$ Balmer series of hydrogen in the HST-1 and HST-2 flares to present new RHD flare models, while also incorporating LTE modeling of the Fe II lines and analysis of the Fe II/C2650\prim\ ratios.  
The Balmer jump ratios must be lower than predicted by models with $T\sim10,000$ K plasma at low continuum optical depth (in the RHD F11 model).  Model atmospheres that produce a dense chromospheric compression with a temperature increase above 10,000 K at moderately high column mass of log $m \sim -2.35$ are expected to produce enough hydrogen Balmer bound-free (photoionization) opacity to lower the Balmer jump ratio to a value of $3-4$ \citep{K18}.  

The electron beam flux regime of $\sim$F12 may help explain the amount of heating at high column mass in these flares, while also favorably resulting in a factor of ten lower return current electric field strength than the multithread F13 model.  In particular, using a high low-energy cutoff causes the beam flux to heat high column mass (log $m \sim -2.35$) to temperatures near $T\sim10^4$ K in order to produce wavelength-dependent optical depths (Paper II), which lowers the Balmer jump ratio from the optically thin, $T=10,000$ K value.  Recent high spatial resolution data of solar flares sometimes imply F12 electron flux densities \citep[L. Fletcher, priv. communication,][]{Fletcher2007, Krucker2011, Sharykin2017}. The timescales and energies of the two GJ 1243 flare events are strikingly similar to the largest solar flares \citep{Woods2004}; thus, HST-1 and HST-2 may be suitable for establishing connections and differences (e.g., in the slow rise and decay of Ca II K) between dMe and dG flares. 

Recently, \citet{Namekata2017} compared the durations and energies of flares on GJ 1243 (from \emph{Kepler}) to white-light flares on the Sun (from SDO/HMI).  They found that both GJ 1243 and the Sun exhibited similar power-law scalings between flare energy and duration as found in \citet{Maehara2015}, but that for a given duration, the flares on GJ 1243 were 10x larger energy (all bolometric flare energies were calculated assuming a blackbody; see Section \ref{sec:discussion3}). The superflares on rapidly rotating G dwarfs were 1000x larger energy for similar durations as solar flares.

We speculate that the \citet{Namekata2017} measure of impulsiveness (location on the duration-energy diagram) may be related to the values of the Balmer jump ratios at peaks of flares and the measures of impulsiveness (from K13) calculated in our work.  However, more measurements of the Balmer jump ratio are needed for the flares of GJ 1243 and for solar flares.  The recent detection and interpretation of NUV continuum radiation in solar flares using IRIS spectra (near $\lambda=2826$ \AA, roughly similar to our measure of C2820) suggests that the flare intensity can be explained by electron beam models that produce optically thin Balmer continuum radiation \citep{Heinzel2014, Kleint2016}.  Even larger Balmer jump ratios are thus expected in solar flares than achieved in the HST-1 and HST-2 events \citep{Kowalski2017A}, but intensity-calibrated spectra with broad wavelength coverage in the $U$-band during solar flares
are very rare \citep{KCF15, Ondrej1}.   Instrumentation in the near future will likely achieve the precision to measure the spectral shape in the $U$-band and NUV during superflares on rapidly rotating dG stars.  

\section{Discussion III:  Broader Implications of NUV Flare Observations} \label{sec:discussion3}

NUV flare spectra are important for input to photochemical modeling of biosignatures and ozone chemistry in planetary atmospheres in the habitable zone of M dwarfs at $\sim0.03-0.1$ a.u.  The nearest Earth-mass planet in the habitable zone was recently discovered around the dM5.5e flare star Proxima Centauri \citep{Anglada2016}, but atmospheric evolution studies are not able to accurately account for the NUV flare irradiation due to a lack of data in this range \citep{Ribas2016, Ribas2017}.  Instead, photochemical modeling studies usually assume that the gradual phase NUV spectra from the Great Flare (a giant IF-type event) from \citet{HP91} represents all flare phases and all flares \citep{Segura2010, Ranjan2017}.  From spectral observations at $\lambda>3500$ \AA, we know that dMe flares exhibit significant inter- and intra-flare variation of the Balmer jump ratio and the blue-to-red optical color temperature (K13, K16).  If 
 exoplanet chemistry models need only very coarse spectral resolution and they have measurements of the Balmer jump and impulsiveness index, the $\lambda<3000$ \AA\  flux can be estimated on one-minute timescales using the relationship that the average flare-only specific flux from $\lambda=2510-2841$ \AA\ is approximately equal to C3615\prim\ for these types of flares (Figure \ref{fig:ultimate}).  

Several smaller dMe flares have been observed with spectra in the NUV.
The NUV was observed with the HST/STIS during several flares (without ground based spectra) on YZ CMi \citep{Hawley2007};  the strength of the continuum flux relative to the emission lines varied from flare to flare:  at $\lambda=2300-3050$ \AA, the percentage of energy in the continuum was typically 70-90\% and 50\% for some smaller events.  The fraction of energy in the NUV continuum in our flare events ($2510-2841$ \AA) is also relatively small $\sim60$\%. 
Thus, a careful consideration is required for using any flare event as a ``template'' for all flares, whether it be for the purposes of photochemistry modeling or comparison to the physical conditions in solar flares. The flare and quiescent spectra in Figure \ref{fig:ultimate} are presented as FITS tables on Zenodo and VizieR, linked to this article.  We hope they will be of use to the exoplanet community as an alternative to the Great Flare data of AD Leo, when HF/GF type events are observed, though the rates of such events are not well-quantified.

Many studies also use a $T \sim 9000$ K blackbody as a fiducial flare spectrum, either to calculate bolometric stellar flare energies from Kepler \citep[e.g.,][]{Maehara2012}, to calculate bolometric solar flare energies \citep{Kretzschmar2011, Namekata2017}, or to calculate an estimate of the ultraviolet flux for exoplanetary atmosphere modeling \citep[e.g.,][]{Howard2018, Loyd2018}.   Our data show that scaling Kepler observations of flares with a $T=9000$ K blackbody to represent the bolometric energy under-estimates the NUV continuum specific flare-only flux at $\lambda < 2840$ \AA\ by a factor of two and underestimates the average $2510-2841$ \AA\ specific flare-only flux by a factor of three.  A $\sim$9000 K blackbody has been found to represent the broadband distribution of several large and moderate energy dMe flares \citep{HF92, Hawley2003}, and this has become a widely assumed property of dM flares.   We convolve the HST-1 impulsive phase spectrum from Figure \ref{fig:fullsed} and Figure \ref{fig:ultimate} with the broadband filters $UBVR$ from \citet{Bessel2013}, and we fit a blackbody to the filter-weighted specific flux densities and a UV continuum point (C2650\prim) following \citet{Hawley2003}.  The best-fit blackbody temperature is 9200 K (Figure \ref{fig:convolve}), which demonstrates that spectra are necessary to conclude that hot blackbody radiation dominates the continuum distribution of dM flares \citep{Allred2006}.  Accurate absolute $U$-band spectrophotometry is subject to several vagaries associated with incomplete filter coverage near the atmospheric limit and differences in total system (telescope$+$atmosphere) response compared to photometry (e.g., in Hawley et al. 2003).   We note that adjusting the $U$-band spectrophotometry in Figure \ref{fig:convolve} by 15\% does not change the inferred blackbody temperature by more than 400 K.  

Figure \ref{fig:convolve} also shows the convolved fluxes through the Large Synoptic Survey Telescope (LSST) $ugri$ filters\footnote{\url{https://github.com/lsst/throughputs/tree/master/baseline}, using the ``total'' filter profiles that are the predicted total system (including atmospheric) response.};  a blackbody with a temperature of $\sim15,000 - 22,000$ K is (poorly) fit without the continuum point in the UV at $\lambda < 3000$ \AA.  This also calls into question the very high blackbody temperatures of $17,000-22,000$ K reported by \citet{Zhilyaev2007} during moderate-amplitude flares on EV Lac using broadband $UBVRI$ photometry.

\begin{figure}
\begin{center}
\includegraphics[scale=0.70]{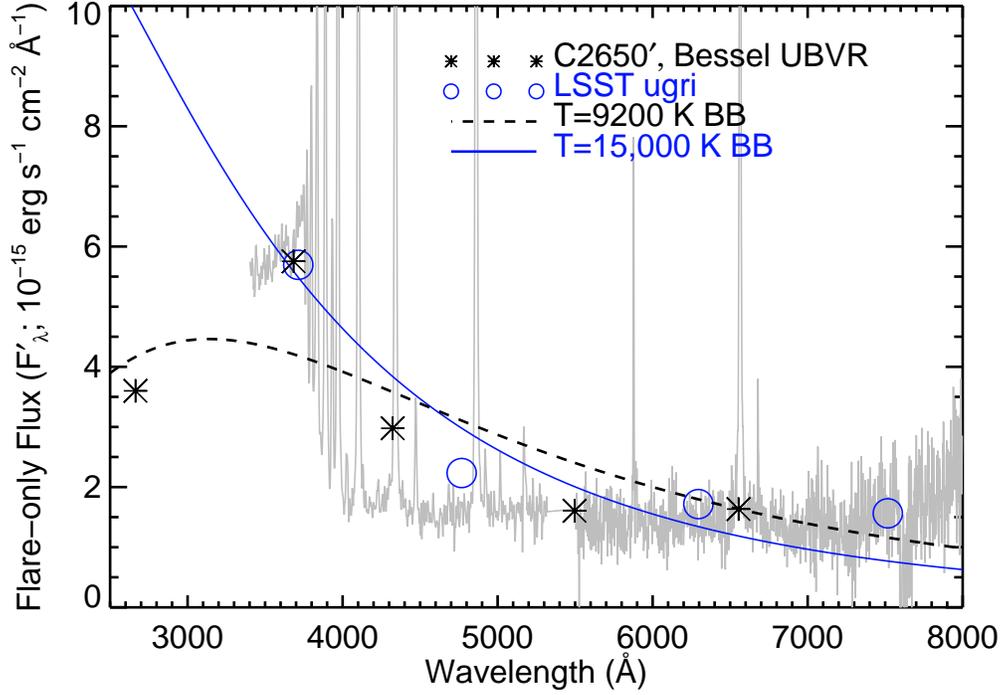}
\caption{HST-1 impulsive phase spectrum (S\#163) from the WHT shown with the C2650\prim\ value from HST/COS (\emph{gray}).  The filter-weighted specific flux densities are shown through the broadband $UBVR$ filters (\emph{asterisks}) and the LSST $ugri$ filters (\emph{open circles}).  The best-fit blackbody function to the C2650\prim$+UBVR$ filters is a dashed black line, and the best-fit blackbody function (with very large uncertainties being so far in the Rayleigh-Jeans tail) to the LSST filters is a \emph{solid blue line}.  The broadband flux distribution of this HF/GF-type event is fit by a blackbody with $T=9200$ K, but the spectrum (\emph{gray}) is dominated by a flat continuum in the red optical and Balmer continuum radiation in the $U$ band and NUV.  }
\end{center}
\end{figure}\label{fig:convolve}

\clearpage
\section{Summary \& Conclusions} \label{sec:conclusions}

We present data from a multi-wavelength flare campaign to characterize the peak of the white-light flare continuum radiation with simultaneous NUV spectra from HST and optical spectra from ground-based telescopes.   We monitored GJ 1243 and observed two events with moderate $U$-band amplitudes ($\Delta U = -1.5$ mags) at peak.  This is the second study to present $\lambda < 3000$ \AA\ NUV flare spectra with simultaneous, flux-calibrated Balmer jump spectra.  Compared to the spectra of a flare event on AU Mic presented in \citet{Robinson1993, Robinson1995}, our data have much higher spectral resolution at $\lambda<3000$ \AA, higher time-resolution by a factor of twenty, and broader spectral continuum characterization to the $U$-band atmospheric limit and into the red and infrared.   In the future, we intend to compare our data to the AU Mic flare, for which there has been little quantitative analysis presented.  

 According to the classification scheme in K13, the photometric and spectral properties of the two events on GJ 1243 observed by HST/COS are most similar to HF/GF-type events, even though the light curves exhibit a fast, impulsive evolution in a by-eye assessment (e.g., Figure \ref{fig:hstlc}).
 The HF/GF characterization means that their spectra exhibit moderately large Balmer jumps and prominent Balmer line radiation; the impulsive phases actually are relatively long and gradual relative to the peak amplitudes.
The goal of our study was to confirm and characterize the extension of the hot $T\sim10^4$ K blackbody-like continuum into the NUV and constrain the white-light peak, but the blue-optical radiation in these two events was rather faint compared to many of the events in K13.

Our conclusions are the following:
\begin{itemize}
\item We detect significant (at signal-to-noise $\ge 20$) broadband NUV continuum increase in the HST spectra.  At relatively low ($\sim60$~s) temporal resolution and cadence, the NUV continuum flux in the HST data follows the continuum time-evolution in the blue optical and the NUV corresponding to the $U$-band.  All continuum fluxes decay faster than the emission lines relative to the respective peak flux of each measure.  Further investigation of this time-decrement with modeling is warranted; however, current RHD simulations exhibit heating for only several seconds whereas these events produce continuum radiation throughout the impulsive phase lasting a few minutes.

\item At high-time resolution ($\sim5$~s cadence), the peak-normalized light curves of the $U$-band and NUV are not identical. Further understanding of this effect requires more high-time resolution NUV data of all flare types (IF, HF, and GF) for robust confirmation that is independent of uncertainties in peak light curve normalization.  From GALEX (photometry), the FUV exhibits faster timescales than the $U$-band \citep{Hawley2003} and the NUV in GALEX \citep{Robinson2005, Welsh2006}.  We speculate that there may be a common physical origin for the shorter peak-normalized decay rates at shorter ultraviolet continuum wavelengths in dMe flares.

\item High-time resolution at 5~s cadence is critical for identifying the fast and slow decay phases of flares (for some IF-type events, even higher time resolution is necessary; K16).  The impulsive phases of the flares of GJ 1243 in the short cadence Kepler data may be unresolved by a factor of ten or more.  Higher time resolution than a 5~s cadence is preferable, since some impulsive phase spikes in the HST flares (e.g., the spike at the beginning of S\#162 in Figure \ref{fig:lcfigs}(a)) appear unresolved.

\item In the NUV, the flare-only spectrum is not a ``scaled-up'' version of the pre-flare spectrum. In other words, there really is significant continuum radiation that appears only during the flare, which is evident from the large variation of the line-to-continuum ratios from quiescence to flaring times.  

\item From spectra, we have determined that continuum flare-only specific flux near $\lambda \sim 2650$ \AA\ is approximately 60\% of the continuum flare-only specific flux in the $U$-band for the two flares studied here. The value of the continuum peak ($\lambda_{\rm{peak}}$) for HF/GF-type events is thus in the $U$-band near the Balmer series limit in these events.  If the pseudo-continuum of merged hydrogen lines just redward of the Balmer limit is due to the dissolved level continuum opacity (to be modeled in detail in Paper II), the value of $\lambda_{\rm{peak}} \sim 3700$ \AA.

\item A $T\sim9000$ K blackbody is a poor approximation to the NUV spectra of HF/GF-type events, even though this blackbody temperature fits the general broadband color distribution and narrowband blue-optical continuum windows.  There is a significant Balmer continuum contribution in the NUV and $U$-band.  A hot blackbody extrapolation from blue-optical wavelengths (longer than the Balmer jump) under-estimates the NUV, $\lambda < 2840$ \AA\ continuum flare-only specific flux by a factor of two.  

Higher spectral resolution at $\lambda=4000-4800$ \AA\ is needed to characterize the blending of many minor emission lines in HF/GF-type events for more robust comparisons to a $T=9000$ K hot blackbody shape in the blue-optical (since the red-optical is rather flat).  The moderate Balmer jump ratios and very large H$\gamma$/C4170\prim\ values from the low-resolution spectra of HST-1 and HST-2 rule out the F11 RHD electron beam heating simulations, but higher spectral resolution data would help confirm that the blue continuum in C4170\prim\ does also not have significant pseudo-continuum of blended emission lines in these (HF/GF) types of events.  

\item K13 showed that the slow rise of Ca II K occurs in IF, HF, and GF-type events.   It is not known why Ca II weakly responds in the dMe flare impulsive phase, but these new spectral constraints of HF events will help provide new insight:  The delayed peak of Ca II K is not related to the presence of spectrally confirmed, energetically dominant hot, blackbody-like radiation.

\item With only broadband photometry, such as in future flare detections with the LSST, HF/GF-type events with moderate Balmer jumps may erroneously suggest very hot blackbody temperatures of $T\sim15,000-20,000$ K.

\item The HF/GF-type events on GJ 1243 should be modeled as inhomogeneously emitting flare sources.    A multithread model approach with F13 beams, previously used to successfully model the decay phase white-light of superflares \citep{Osten2016, Kowalski2017B}, can adequately explain the continuum flux distribution from $2500-2840$ \AA\ and from $\sim3200-4200$ \AA.

\end{itemize}

The HST-1 and HST-2 events establish a new regime of flare-only continuum flux colors over the impulsive phase of dMe flares with moderate Balmer jump ratios ($3-4$) and relatively flat blue-to-red optical continuum shapes (blue-to-red continuum flux ratios of $1-1.4$).  Other moderate-amplitude events, such as the IF4 and IF11 events on YZ CMi from K16 (also discussed in Section \ref{sec:discussion1} here) exhibit significantly different continuum and line-to-continuum ratios calculated from spectra at a similar temporal resolution ($t_{\rm{exp}}\sim60$~s).  The value of $\lambda_{\rm{peak}}$ for IF-type events with smaller Balmer jump ratios is a critical parameter for a comprehensive understanding of the heating at high column mass achieved in dMe flares.  IF-type events have shorter impulsive phase timescales (and sometimes much more luminous peaks).  Even with low-to-moderate signal-to-noise, a strategic use of COS/G230L based on our results could provide representative constraints on $\lambda_{\rm{peak}}$ in the impulsive phase of giant impulsive-type events, such as the Great Flare of AD Leo.

\acknowledgments
We thank an anonymous referee for a critical reading of the paper and helpful comments and suggestions.  
AFK acknowledges support from University of Maryland GPHI Task-132, HST GO 13323, NASA Exoplanet Science Institute (NASA/Keck time), an appointment to the NASA Postdoctoral Program at the NASA's Goddard Space Flight Center, administered by Universities Space Research Association (previously by the Oak Ridge Associated Universities) under contract with NASA.  AFK thanks Joel Allred and Mats Carlsson for many helpful discussions about RADYN, Han Uitenbroek for helpful discussions about the RH code, and Pier-Emmanuel Tremblay for the hydrogen broadening profiles.  AFK thanks James Davenport for observations from the ARCSAT 0.5-m and for providing a flare rate for GJ 1243, Nicola Gentile Fusillo for observations from the Isaac Newton Telescope, Mihalis Mathioudakis for helpful discussions about the ULTRACAM data, Lyndsay Fletcher for helpful discussions about the interpretation of the HST data, Lucianne Walkowicz for helpful feedback on the initial HST proposal, and Lucia Kleint for helpful discussions on comparing to blackbodies.  We thank R. O. Parke Loyd and the STScI help desk for helpful feedback on the jump in the dispersion solution in the COS spectra.  
AFK also thanks the Keck Observatory support astronomer Luca Rizzi for assistance with the observations.  
IRAF is distributed by the National Optical Astronomy Observatory, which is operated by the Association of Universities for Research in Astronomy (AURA) under a cooperative agreement with the National Science Foundation.
Some of the data presented in this paper were obtained from the Mikulski Archive for Space Telescopes (MAST). STScI is operated by the Association of Universities for Research in Astronomy, Inc., under NASA contract NAS5-26555.  Support for this work was provided by NASA through grant number Guest Observer 13323 from the Space Telescope Science Institute, which is operated by AURA, Inc., under NASA contract NAS 5-26555.  
Based on observations obtained with the Apache Point Observatory 3.5-m telescope, which is owned and operated by the Astrophysical Research Consortium.
The 2.3-m Aristarchos telescope is operated on Helmos Observatory by the Institute for 
Astronomy, Astrophysics, Space Applications and Remote Sensing of the National
Observatory of Athens.
This work has made use of data from the European Space Agency (ESA) mission
{\it Gaia} (\url{https://www.cosmos.esa.int/gaia}), processed by the {\it Gaia}
Data Processing and Analysis Consortium (DPAC,
\url{https://www.cosmos.esa.int/web/gaia/dpac/consortium}). Funding for the DPAC
has been provided by national institutions, in particular the institutions
participating in the {\it Gaia} Multilateral Agreement.

\end{document}